\documentclass[prd,notitlepage,longbibliography,superscriptaddress,nofootinbib]{revtex4-2} 

\usepackage{amsmath}
\usepackage{amssymb}
\usepackage{bm}
\usepackage{dsfont}
\usepackage{graphicx}
\usepackage{xcolor}
\usepackage{array}
\usepackage{cleveref}
\usepackage{tabularx}
\usepackage{mathrsfs} 
\usepackage{mathtools}
\usepackage{booktabs}
\usepackage{fancyhdr}
\usepackage{float}
\usepackage{tabularx}
\usepackage{capt-of}
\usepackage{adjustbox}

\DeclareMathOperator{\Tr}{Tr}

\begin{document}
\preprint{}
\title{Scalar quasi-normal modes of accelerating Kerr-Newman-AdS black holes}

\author{Juli\'{a}n Barrag\'{a}n Amado}
  \email{barragan86@dongguk.edu}
  \affiliation{Division of Physics and Semiconductor Science, Dongguk University, Seoul 04620, Republic of Korea}
\author{Bogeun Gwak}
  \email{rasenis@dgu.ac.kr}
  \affiliation{Division of Physics and Semiconductor Science, Dongguk University, Seoul 04620, Republic of Korea}

\begin{abstract}
We study linear scalar perturbations of slowly accelerating Kerr-Newman-anti-de Sitter black holes using the method of isomonodromic deformations. The conformally coupled Klein-Gordon equation separates into two second-order ordinary differential equations with five singularities. Nevertheless, the angular equation can be transformed into a Heun equation, for which we provide an asymptotic expansion for the angular eigenvalues in the small acceleration and rotation limit. In the radial case, we recast the boundary value problem in terms of a set of initial conditions for the isomonodromic tau function of Fuchsian systems with five regular singular points. For the sake of illustration, we compute the quasi-normal modes frequencies.
\end{abstract}

\keywords{Quasinormal modes, Accelerating Kerr-Newman-AdS Black Hole, Isomonodromic tau function}


\maketitle

\section{Introduction}
\label{sec:1}

The anisotropic emission of gravitational waves from the merger of black hole (BH) binaries suggests that the final remnant recoils from the center-of-mass frame with a velocity that depends on the configuration of the system and the details of the merger dynamics but not on the total mass  (see \cite{Bruegmann:2007bri} and references therein). Namely, large recoils may exceed the escape velocity, and consequently, BHs can get ejected from their host galaxies \cite{Gerosa:2016vip}. This effect, known as BH kick, requires a better understanding of moving and accelerating black holes. The Pleba\'{n}ski-Demia\'{n}ski family of solutions includes the $C$-metric, which describes two uniformly accelerated black holes in opposite directions and can be generalized to contain rotation, charge, and cosmological constant. Therefore, it stands as a natural candidate to study boosted black holes \cite{Griffiths:2005se,Krtous:2005ej,Podolsky:2006px}. 
In this context, numerical and analytical studies have addressed the computation of the QNMs, the character of the black hole shadow, and the stability of accelerating black holes \cite{Destounis:2020pjk,Destounis:2020yav,Fontana:2022whx,Destounis:2022rpk,Gwak:2022nsi,Zhang:2023yco,Xiong:2023usm,Lei:2023mqx,SUI2023138135}.

In general, the global structure of the $C$-metric in flat, de Sitter or anti-de Sitter represents two accelerated black holes. The acceleration $A$ is caused by the conical singularities located along the axis of symmetry and can be thought of as strings pulling the two black holes apart, or a strut between them. 
Furthermore, it is responsible for the appearance of an acceleration horizon between the outer horizon and the spatial infinity, which possesses the same causal structure of the cosmological horizon of the de Sitter space-time. However, in asymptotically AdS the relation between the acceleration $A$ of the black holes and the cosmological length $L$ gives rise to three possibilities: 1) For $A < 1/L$, the metric describes one accelerated black hole in AdS. 2) For $A > 1/L$, it describes a pair of causally separated accelerated black holes. 3) In the critical case $A = 1/L$, one has one single accelerated black hole entering and leaving asymptotically AdS \cite{Dias:2002mi}.

Despite the interest on these solutions, the thermodynamics of accelerating black holes remained less well understood. Recently, a consistent black hole thermodynamics has been formulated in the slowly accelerating black hole limit \cite{Anabalon:2018ydc,Anabalon:2018qfv,Cassani:2021dwa}; see \cite{Kim:2023ncn} for a derivation using the covariant phase space formalism. In this limit, the accelerating black hole does not possess an acceleration horizon, and hence the acceleration of the black hole merely plays the role of an independent parameter.

In this paper, we compute the quasi-normal modes (QNMs) of massless scalar fields on a slowly accelerating Kerr-Newman-anti-de Sitter (KNAdS$_{4}$) black hole using the method of isomonodromic deformations. After separating the conformally coupled Klein-Gordon equation, we transform the radial (angular) ODEs into a Heun-like equation. The resulting equation possesses five regular singular points and can be reduced to the Heun equation. Recently, in the context of conformal mapping of polycircular arc domains, the solution of the direct Riemmann-Hilbert problem for ODEs with $n$-regular singularities has been constructed to find the accessory parameters and the conformal moduli given the monodromy data. Namely, they have provided some examples in the case of $n = 5$ vertices \cite{CarneirodaCunha:2021jsu}. We will be interested in the inverse Riemann-Hilbert problem associated with a Fuchsian system with five regular singular points. Interestingly, the initial conditions that allow us to recast the boundary value problem are the same. The initial conditions are written in terms of an isomonodromic tau function defined in \cite{Gavrylenko:2016zlf} and then reviewed in \cite{CarneirodaCunha:2021jsu} for the particular case of five vertices. The latter approach has been applied in the case of four regular singular points and three singularities - two regular and one irregular singular point, where the isomonodromic deformation equations are related to the celebrated Painlev\'{e} VI and Painlev\'{e} V equations, respectively. As examples, we mention the computation of QNMs \cite{CarneirodaCunha:2015qln,Novaes:2018fry,BarraganAmado:2018zpa,BarraganAmado:2021uyw,CarneirodaCunha:2019tia,Cavalcante:2021scq,Amado:2021erf}\footnote{See also \cite{Aminov:2020yma,Bonelli:2021uvf,Ikeda:2021uvc,Bianchi:2021mft} for another approach based on four-dimensional $\mathcal{N}=2$ supersymmetric gauge theories.}, the Rabi model in quantum optics \cite{CarneirodaCunha:2015vxu} and conformal maps \cite{Anselmo:2018zre,Anselmo:2020bmt}. These efforts have been inspired by the seminal works \cite{Gamayun:2012ma,Gamayun:2013auu}, where the expansion of the isomonodromic tau function was derived in terms of $c = 1$ Virasoro conformal blocks. To our knowledge, this manuscript represents the first study of quasi-normal modes via the isomonodromic tau function of Fuchsian systems with five regular singular points.

It is worth mentioning that our work enhance previous analysis presented in \cite{Wei:2021bqq,Fontana:2022whx}. In the first manuscript, authors found a master equation for massless fields of spin $s = \lbrace 0, 1/2, 1, 3/2, 2 \rbrace$, using the Newman-Penrose formalism \cite{Bini:2008mzd}. Then, by considering the existence of the acceleration horizon in accelerating Kerr-Newman-AdS$_{4}$ black holes, they obtained a solution of the radial equation near the event horizon, while the second article investigates the QNMs of scalar fields propagating on slowly accelerating Reissner-Nordstr\"{o}m-AdS$_{4}$ black holes. In this regard, we have computed the eigenfrequencies for scalar perturbations on a slowly accelerating KNAdS$_{4}$ and provided an asymptotic expansion for the separation constant which in the limit $a \rightarrow 0$ reproduces the results in \cite{Fontana:2022whx}. 

This manuscript is organized as follows. In Section \ref{sec:2}, we introduce the geometry and analyze the space of parameters of the accelerating Kerr-Newman-anti de Sitter black hole. Then, we review the Klein-Gordon equation for massless charged scalar perturbations and separate it into two second-order ODEs. Section \ref{sec:3} is devoted to the analysis of the separation constant and the quasi-normal modes. Namely, we obtain an asymptotic expansion for the angular eigenvalues in the small acceleration and rotation limit via the Painlev\'{e} VI tau function in Subsection \ref{sec:3a}. Subsequently, we compute the QNMs by solving a set of transcendental equations expressed in terms of an isomonodromic tau function in Subsection \ref{sec:3b}. We conclude with a short discussion of the results and future perspectives in Section \ref{sec:4}. 

\section{Accelerating Kerr-Newman black holes in AdS}
\label{sec:2}

We consider an accelerating black hole solution derived from the Pleba\'{n}ski-Demia\'{n}sky metric \cite{Plebanski:1976gy} which describes an accelerating and rotating charged black hole in asymptotically anti-de Sitter space-time. In Boyer-Lindquist type coordinates, the line element reads

\begin{subequations}\label{eq:metric}
\begin{equation}\label{eq:line_element}
ds^2=\frac{1}{\Omega^2}\biggl\lbrace -\frac{\Delta_{r}}{\rho^{2}}\left[\frac{dt}{\alpha}-a\sin^{2}\theta\frac{d\varphi}{K}\right]^{2}+\frac{\rho^{2}}{\Delta_{r}}dr^{2}
+\frac{\rho^{2}}{\Delta_{\theta}}d\theta^{2} + \frac{\Delta_{\theta}\sin^{2}\theta}{\rho^{2}}\left[\frac{a\,dt}{\alpha}-(r^{2}+a^{2})\frac{d\varphi}{K}\right]^{2}\biggr\rbrace,
\end{equation}

where the vector potential is

\begin{equation}\label{eq:gauge_field}
B = -\frac{e\,r}{\rho^{2}}\left[\frac{dt}{\alpha}-a\sin^{2}\theta\frac{d\varphi}{K}\right] + \Phi_{t}dt, \qquad F = dB, 
\end{equation}
\end{subequations}

and $\Phi_{t}$ is chosen to vanish the scalar potential at the outer horizon. The metric functions are defined in \cite{Anabalon:2018qfv} as follows

\begin{equation}
\begin{split}
&\Delta_{r}=(1-A^{2}r^{2})\left(r^{2}-2M\,r+(a^{2}+e^{2})\right) + \frac{r^{2}(r^{2}+a^{2})}{L^{2}},\\
&\Delta_{\theta}=1+2MA\cos\theta+\left[A^{2}(a^{2}+e^{2})-\frac{a^{2}}{L^{2}}\right]\cos^{2}\theta\\
&\rho^{2}=r^{2}+a^{2}\cos^{2}\theta, \qquad \Omega=1+Ar\cos\theta,
\end{split}
\end{equation}

where $M,a,e,L$ are related to the mass, the rotation parameter, the charge of the black holes and the radius of AdS, respectively. In addition, $A$ represents the acceleration of the black holes, while $K$ fixes the range of the azimuthal coordinate $\varphi$ by removing one of the conical singularities and the factor $\alpha$ is a scaling factor that normalizes the time-like Killing vector to obtain the correct thermodynamics \citep{Anabalon:2018ydc,Anabalon:2018qfv}
\begin{equation}\label{eq:alpha}
\alpha = \frac{\sqrt{\left(\Xi + \frac{a^{2}}{L^{2}}\right)\left(1 - A^{2}L^{2}\Xi\right)}}{1+a^{2}A^{2}}, \qquad \Xi = 1-\frac{a^{2}}{L^{2}} + A^{2}\left(e^{2} + a^{2}\right).
\end{equation}

The function $\Delta_{r}$ is a quartic function of the radial coordinate\footnote{For $A=1/L$, $\Delta_{r}$ reduces from a quartic to a cubic function in $r$ (see \cite{Emparan:1999wa} for a detailed studied on this limit).}, where the black hole horizons correspond to the roots of $\Delta_{r}$, and can be written as
\begin{equation}\label{eq:deltar}
\Delta_{r} = \dfrac{\left(1-A^{2}L^{2}\right)}{L^{2}}\left(r-r_{0}\right)\left(r-r_{1}\right)\left(r-r_{2}\right)\left(r-r_{3}\right).
\end{equation}
Moreover, since we are interested in space-time configurations that only possess black hole horizons, the event horizon will correspond to the largest positive real root $r_{3} = r_{+}$ in $\Delta_{r}$, while the Cauchy horizon is related to $r_{2} = r_{-}$ satisfying $ r_{-} < r_{+}$, yielding
\begin{equation}\label{eq:slowly_deltar}
\Delta_{r} = \dfrac{\left(1-A^{2}L^{2}\right)}{L^{2}}\left(r-r_{0}\right)\left(r-r_{1}\right)\left(r-r_{-}\right)\left(r-r_{+}\right).
\end{equation}

The latter assumption is constrained by the following conditions on the metric \eqref{eq:metric}: (i) $\Delta_{\theta} > 0$ in the range $\left[0, \pi\right]$, which implies

\begin{equation}\label{eq:black_curve}
M A <
\begin{cases}
\frac{1}{2}\Xi,&  \Xi \in  \left(0, 2\right], \\
\\
\sqrt{1 - \Xi},& \Xi > 2,
\end{cases}
\end{equation}

(ii) the existence of a black hole within the boundaries of the space-time, which can be thought of as $\Delta_{r}(r_{+}) = 0 = \Delta_{r}^{\prime}(r_{+)} $ with at least two roots in the range $r \in \left(0, 1/A\right)$. (iii) The absence of acceleration and cosmological horizons in the bulk. Black holes satisfying the latter conditions have been referred as slowly accelerating Kerr-Newman-AdS$_{4}$ black holes \citep{Anabalon:2018qfv}. From the thermodynamics point of view the temperature of the slowly accelerating KN-AdS$_{4}$ black hole at the event horizon $r_{+}$ is given by

\begin{equation}\label{eq:temperature}
T_{+} = \frac{1}{4\pi\alpha}\frac{\Delta_{r}^{\prime}(r_{+})}{r_{+}^{2}+a^{2}}.
\end{equation}

Following the study of the parameter space for slowly accelerating Reissner-Nordstr\"{o}m black holes presented in \cite{Abbasvandi:2019vfz}, we introduce the dimensionless parameters

\begin{equation}\label{eq:parameters}
\tilde{A} = A L, \qquad \tilde{M} = M A, \qquad \tilde{a} = a A, \qquad \tilde{e} = e A,
\end{equation}

to explore the admissible parameter space in the dimensionless $\bigl(\tilde{A}, \tilde{M}, \tilde{a}, \tilde{e}\bigr)$ plane that describes slowly accelerating Kerr-Newman-AdS black holes configurations (see \cite{Anabalon:2018ydc,Anabalon:2018qfv,EslamPanah:2019szt,Cassani:2021dwa,Kim:2023ncn} for a detailed discussion on the role of conical deficits in the thermodynamics of accelerating black holes in AdS).

\begin{figure}[!tbh]
\centering
\includegraphics[width=0.85\linewidth]{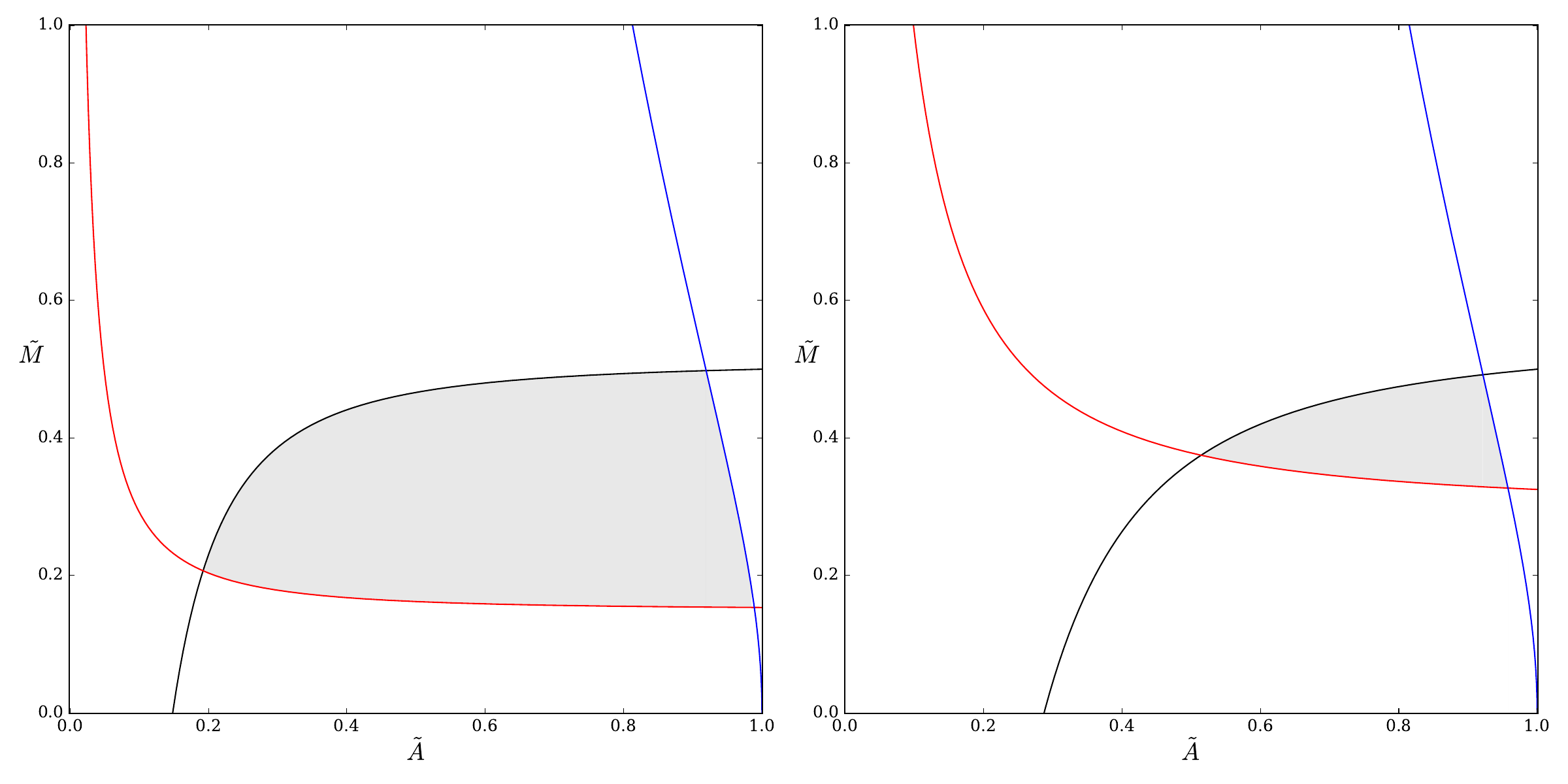}
\caption{The parameter space of a slowly accelerating Kerr-Newman-AdS$_{4}$ black hole is defined as the region formed by the intersection of the three curves in the $\tilde{A}-\tilde{M}$ plane. The left and right figures exhibit the parameter space for $\left( \tilde{a}_{L} = 0.15,\, \tilde{e}_{L} = 0.0001 \right)$ and $\left( \tilde{a}_{R} = 2\tilde{a}_{L},\, \tilde{e}_{R} = 2\tilde{e}_{L} \right)$, respectively. Furthermore, it is worth noting that the area resulting from the intersection of the three boundary curves (namely, black, red, and blue) decreases as the values of $\tilde{a}$ and $\tilde{e}$ increase.}
\label{fig:space}
\end{figure}

For fixed rotation and charge values, we display in the $\tilde{A}-\tilde{M}$ plane the curves that correspond to these constraints. In Figure \ref{fig:space}, we show the parameter space of a slowly accelerating KN-AdS$_{4}$ for fixed $\left(\tilde{a}, \tilde{e}\right)$ values. The first condition \eqref{eq:black_curve} is satisfied for black hole configurations under the black curve, while the second condition is guaranteed by black holes parameters located over the red curve. The last condition prevents the formation of acceleration horizons for configurations to the left of the blue curve, or in other words, spacetimes with additional horizons are to the right of this curve. It is worth mentioning that $\tilde{A}^{2} < 1$ is not a sufficient condition for slowly accelerating black holes since there are values of $\tilde{M}$ outside the shaded region even though the aforementioned inequality is satisfied.

\subsection{The Klein-Gordon equation}
\label{sec:2.1}
In this section we will consider the conformally coupled Klein-Gordon equation on the metric \eqref{eq:metric} for massless and charged scalar fields
\begin{equation}\label{eq:klein_gordon}
\left[\frac{1}{\sqrt{-g}}D_{\mu}\left(\sqrt{-g}g^{\mu\nu}D_{\nu}\right) - \xi R\right]\Psi = 0,
\end{equation}
where $D_{\mu} = \nabla_{\mu} - i q B_{\mu}$, $q$ is the charge of the field, and $\xi = 1/6$ is the coupling constant in four-dimensions. The equation \eqref{eq:klein_gordon} is not separable when using the metric \eqref{eq:metric} due to the conformal factor $\Omega$. However, by performing the following conformal transformation
\begin{equation}\label{eq:conformal_trans}
\tilde{g}_{\mu\nu} = \Omega^{2}g_{\mu\nu},
\end{equation}
one can get the conformal metric
\begin{equation}\label{eq:conformal_metric}
d\tilde{s}^2= -\frac{\Delta_{r}}{\rho^{2}}\left[\frac{dt}{\alpha}-a\sin^{2}\theta\frac{d\varphi}{K}\right]^{2}+\frac{\rho^{2}}{\Delta_{r}}dr^{2}
+\frac{\rho^{2}}{\Delta_{\theta}}d\theta^{2} + \frac{\Delta_{\theta}\sin^{2}\theta}{\rho^{2}}\left[\frac{a\,dt}{\alpha}-(r^{2}+a^{2})\frac{d\varphi}{K}\right]^{2},
\end{equation}
while the scalar field transforms as $\tilde{\Psi} = \Omega^{-1}\Psi$ in four dimensions (we refer to Appendix D in \cite{Wald:1984rg} for further details). The latter transformations ensure the separability of the scalar wave equation
\begin{equation}\label{eq:conformal_equation}
\left[\frac{1}{\sqrt{-\tilde{g}}}\tilde{D}_{\mu}\left(\sqrt{-\tilde{g}}\tilde{g}^{\mu\nu}\tilde{D}_{\nu}\right) - \frac{1}{6} \tilde{R}\right]\tilde{\Psi} = 0,
\end{equation}
which in the conformally rescaled metric \eqref{eq:conformal_metric} reduces to
\begin{equation}
\begin{split}
\biggl\lbrace\frac{\partial}{\partial r}\left(\Delta_{r}\frac{\partial}{\partial r}\right) &+ \frac{1}{\sin\theta}\frac{\partial}{\partial \theta}\left(\sin\theta \Delta_{\theta}\frac{\partial}{\partial \theta}\right) + \frac{\alpha^{2}(r^{2}+a^{2})^{2}}{\Delta_{r}}\left[i\frac{\partial}{\partial t}+q\Phi_{t} - \frac{q\,e\,r}{\alpha(r^{2}+a^{2})} + i\frac{a\,K}{\alpha(r^{2}+a^{2})}\frac{\partial}{\partial \varphi}\right]^{2}\\
&+\frac{\sin^{2}\theta}{\Delta_{\theta}}\left[\alpha\,a\left(\frac{\partial}{\partial t} - i\,q\Phi_{t}\right) + \frac{K}{\sin^{2}\theta}\frac{\partial}{\partial \varphi}\right]^{2} + \frac{1}{6} \left(\Delta_{r}^{\prime\prime} + \Delta_{\theta}^{\prime\prime}+3\cot\theta\Delta_{\theta}^{\prime}-2\Delta_{\theta}\right)\biggr\rbrace\tilde{\Psi}=0.
\end{split}
\end{equation}
Then, by defining the following Ansatz
\begin{equation}\label{eq:ansatz}
\tilde{\Psi}(t,r,\theta,\varphi) = e^{-i\omega t + i m\varphi}R(r)S(\theta),
\end{equation}
we obtain two second-order ODEs of the form
\begin{subequations}\label{eq:ODEs}
\begin{equation}\label{eq:angular_ode}
\begin{split}
\frac{d}{d\theta}\left(\sin\theta\Delta_{\theta}\frac{d S}{d\theta}\right) + \biggl\lbrace &-\frac{\sin^{3}\theta}{\Delta_{\theta}}\biggl[\alpha\,a\left(\omega+q\Phi_{t}\right)-\frac{K\,m}{\sin^{2}\theta}\biggr]^{2}\\
&\qquad + \frac{1}{6}\left(\sin\theta\Delta_{\theta}^{\prime\prime}+3\cos\theta\Delta_{\theta}^{\prime}-2\sin\theta\Delta_{\theta}\right) +\lambda\sin\theta\biggr\rbrace S(\theta) = 0,
\end{split}
\end{equation}
\begin{equation}\label{eq:radial_ode}
\begin{split}
\frac{d}{d r}\left(\Delta_{r}\frac{d R}{d r}\right) + \biggl\lbrace\frac{\alpha^{2}(r^{2}+a^{2})^{2}}{\Delta_{r}}\left[\omega+q\Phi_{t}-\frac{q\,e\,r}{\alpha(r^{2}+a^{2})}-\frac{a\,K\,m}{\alpha(r^{2}+a^{2})}\right]^{2} + \frac{1}{6}\Delta_{r}^{\prime\prime}-\lambda\biggr\rbrace R(r)=0,
\end{split}
\end{equation}
\end{subequations}
where $\omega$ is the mode frequency, $m$ is the (magnetic) azimuthal quantum number, and $\lambda$ is the separation constant associated with the angular eigenvalue equation. These equations \eqref{eq:ODEs} are a particular case $\left(p = 0\right)$ of those derived in \cite{Wei:2021bqq}. 

\subsubsection{The Angular Equation}
\label{sec:2.1.1}

By defining  $u=\cos\theta$, the angular equation \eqref{eq:angular_ode} can be rewritten as
\begin{equation}\label{eq:angular_odeS}
\begin{split}
\frac{d^{2} S}{d u^{2}} + \biggl[\frac{1}{u+1}+\frac{1}{u-1}+\frac{1}{u-u_{0}}+\frac{1}{u-u_{1}}\biggr]\frac{d S}{d u} &+ \biggl\lbrace -\frac{1}{\Delta_{u}^{2}}\left(\alpha\,a(\omega+q\Phi_{t}) - \frac{K\,m}{(1-u^{2})}\right)^{2} \\ 
&+ \frac{1}{6}\biggl(\frac{\Delta_{u}^{\prime\prime}}{\Delta_{u}} - \frac{4 u \Delta_{u}^{\prime}}{(1-u^{2})\Delta_{u}} - \frac{2}{(1-u^{2})}\biggr) + \frac{\lambda}{(1-u^{2})\Delta_{u}}\biggr\rbrace S(u) = 0,
\end{split}
\end{equation}
with
\begin{equation}
\Delta_{u} = \gamma\left(u-u_{0}\right)\left(u-u_{1}\right), \qquad \gamma= \left(A^{2}e^{2}-\frac{a^{2}}{L^{2}}(1-A^{2}L^{2})\right).
\end{equation}
This equation possesses five singular points located at $u = \lbrace -1, 1, u_{0}, u_{1}, \infty \rbrace$, where the roots of the indicial equation, associated to the Frobenius solutions near each singular point, are given by
\begin{equation}
\beta^{\pm}_{-1} = \pm\frac{1}{2}\rho_{-1}, \quad \beta^{\pm}_{1} = \pm\frac{1}{2}\rho_{1}, \quad \beta^{\pm}_{u_0} = \pm\frac{1}{2}\rho_{u_0}, \quad \beta^{\pm}_{u_1} = \pm\frac{1}{2}\rho_{u_1}, \quad \beta^{\pm}_{\infty} = \frac{1}{2}\left(3 \pm 1\right),
\end{equation}
\begin{equation}
\begin{split}
&\rho_{-1} = \frac{K\,m}{\gamma\left(1+u_{0}\right)\left(1+u_{1}\right)}, \qquad \rho_{1} = \frac{K\,m}{\gamma\left(1-u_{0}\right)\left(1-u_{1}\right)},\\
&\rho_{u_{0}} = \frac{2}{\gamma\left(u_{0}-u_{1}\right)}\left(\alpha\,a\left(\omega+q\Phi_{t}\right)-\frac{K\,m}{\left(1+u_{0}\right)\left(1-u_{0}\right)}\right),\\
&\rho_{u_{1}} = \frac{2}{\gamma\left(u_{0}-u_{1}\right)}\left(\alpha\,a\left(\omega+q\Phi_{t}\right)-\frac{K\,m}{\left(1+u_{1}\right)\left(1-u_{1}\right)}\right).
\end{split}
\end{equation}
In order to remove the conical singularity along $\theta = \pi$, we fix $K = \gamma\left(1+u_{0}\right)\left(1+u_{1}\right)$. Furthermore, by introducing a new change of variables
\begin{equation}
w = \frac{u_{0}-u_{1}}{u_{0}+1}\frac{u+1}{u-u_{1}},
\end{equation}
where the conformal moduli are
\begin{equation}\label{eq:ang_conformal_moduli}
w_{1} = \frac{2\left(u_{0}-u_{1}\right)}{\left(1+u_{0}\right)\left(1-u_{1}\right)}, \qquad w_{2} = \frac{u_{0}-u_{1}}{u_{0}+1},
\end{equation}
and a s-homotopic transformation
\begin{equation}\label{eq:shomotopic_angular}
S(w) = w^{\beta^{-}_{-1}}\left(w-w_{1}\right)^{\beta^{-}_{1}}\left(w-w_{2}\right)^{\beta^{-}_{\infty}}\left(w-1\right)^{\beta^{-}_{u_{0}}}y(w),
\end{equation}
we can bring \eqref{eq:angular_odeS} into the canonical form of the Heun equation as the singularity at $w = w_{2}$ turns into a removable singularity \cite{Suzuki:1998vy}\footnote{In Section \ref{sec:3b}, we will use a M\"{o}bius transformation that is more convenient for the computation of the separation constant.}. Namely, we get
\begin{equation}\label{eq:heun_ang_conformal}
\frac{d^{2}y}{d w^{2}} + \biggl(\frac{1 - \rho_{-1}}{w} + \frac{1 - \rho_{1}}{w-w_{1}} +\frac{1 - \rho_{u_{0}}}{w - 1}\biggr)\frac{d y}{d w} + \biggl[\frac{q_{1}q_{2}}{w(w-1)} - \frac{w_{1}(w_{1}-1)Q_{1}}{w(w-w_{1})(w-1)}\biggr]y(w) = 0,
\end{equation}
where
\begin{subequations}
\begin{equation}\label{eq:new_qius}
q_{1} = \dfrac{1}{2}\left(\rho_{-1} + \rho_{1} + \rho_{u_{0}} - 2 - \rho_{u_{1}}\right), \qquad q_{2}=\dfrac{1}{2}\left(\rho_{-1} + \rho_{1} + \rho_{u_{0}} - 2 + \rho_{u_{1}}\right),
\end{equation}
\begin{equation}\label{eq:new_accessory}
\begin{gathered}
Q_{1} = \frac{\hat{\lambda}}{w_{1}(w_{1}-1)} - \frac{1}{w_{1}}\biggl(\frac{1}{3} + \rho_{-1}\rho_{1} - \frac{1}{2}\left(\rho_{-1} + \rho_{1}\right)\biggr) - \frac{1}{w_{1}-1}\biggl( \frac{1}{3} - \frac{1}{2}\left(\rho_{u_{0}} + \rho_{1}\right)\biggr).
\end{gathered}
\end{equation}
\end{subequations}
and the separation constant is given by
\begin{equation}\label{eq:lambdahat}
\hat{\lambda} = \frac{\lambda}{\gamma\left(1+u_{0}\right)\left(1-u_{1}\right)}.
\end{equation}

\subsubsection{The Radial Equation}
\label{sec:2.1.2}
Equation \eqref{eq:radial_ode} possesses five singularities located at the root of $\Delta_{r}$ and the point at infinity. The characteristic exponents of the Frobenius solutions near to each finite singular point are defined by
\begin{equation}
\beta^{\pm}_{k} = \pm\frac{1}{2}\theta_{k}, \qquad k=0,1,-,+, 
\end{equation}
and for $r=\infty$, we have
\begin{equation}
\beta^{\pm}_{\infty}=\frac{1}{2}\left(3 \pm 1\right),
\end{equation}
where
\begin{equation}\label{eq:thetas}
\theta_{k} = \frac{i}{2\pi T_{k}}\left(\omega+q\Phi_{t}-\frac{q\,e\,r_{k}}{\alpha(r_{k}^{2}+a^{2})}-\frac{a\,K\,m}{\alpha(r_{k}^{2}+a^{2})}\right).
\end{equation}
Furthermore, the radial equation can be written on the Riemann sphere by introducing a M\"{o}bius transformation
\begin{equation}\label{eq:Mobius}
z = \frac{r - r_{0}}{r_{+} - r_{0}},
\end{equation}
which fixes three singularities at $z = \lbrace 0, 1, \infty \rbrace$, and the two remaining singular points are situated at
\begin{equation}\label{eq:rad_conformal_moduli}
z_{1} = \frac{r_{1} - r_{0}}{r_{+} - r_{0}}, \qquad z_{2}=\frac{r_{-} - r_{0}}{r_{+} - r_{0}},
\end{equation}
followed by a s-homotopic transformation
\begin{equation}\label{eq:shomotopic_radial}
R(z) = z^{\beta^{-}_{0}} (z - z_{1})^{\beta^{-}_{1}} (z - z_{2})^{\beta^{-}_{-}} (z - 1)^{\beta^{-}_{+}} f(z),
\end{equation}
which brings equation \eqref{eq:radial_ode} into the following form
\begin{equation}\label{eq:heun_radial}
\begin{split}
\frac{d^{2}f}{d z^{2}} + \biggl( \frac{1 - \theta_{0}}{z} + \frac{1 - \theta_{1}}{z - z_{1}} &+ \frac{1 - \theta_{-}}{z - z_{2}} + \frac{1 - \theta_{+}}{z - 1}\biggr)\frac{d f}{d z}\\
&+ \biggl[\frac{\kappa_{1}\kappa_{2}}{z(z-1)} - \frac{z_{1}(z_{1} - 1)K_{1}}{z(z - z_{1})(z - 1)} - \frac{z_{2}(z_{2} - 1)K_{2}}{z(z - z_{2})(z - 1)}\biggr]f(z) = 0,
\end{split}
\end{equation}
where
\begin{subequations}
\begin{equation}\label{eq:kappas}
\kappa_{1} = \dfrac{1}{2}\left(4 - \theta_{0} - \theta_{1} - \theta_{-} - \theta_{+} \right), \qquad \kappa_{2}=\dfrac{1}{2}\left(2 - \theta_{0} - \theta_{1} - \theta_{-} - \theta_{+}\right),
\end{equation}
\begin{equation}\label{eq:accessory1_radial}
K_{1} = -\frac{\tilde{\lambda}}{z_{1}(z_{1} - 1)} - \frac{1}{z_{1}}\left(\frac{1}{3} - \frac{1}{2}\left(\theta_{0} + \theta_{1}\right)\right) - \frac{1}{z_{1} - 1}\biggl(\frac{1}{3} - \frac{1}{2}(\theta_{1} + \theta_{+})\biggr) 
- \frac{1}{z_{1} - z_{2}}\biggl( \frac{1}{3} - \frac{1}{2}(\theta_{1} + \theta_{-})\biggr),
\end{equation}
\begin{equation}\label{eq:accessory2_radial}
K_{2} = \frac{\tilde{\lambda}}{z_{2}(z_{2} - 1)} - \frac{1}{z_{2}}\left(\frac{1}{3} - \frac{1}{2}\left(\theta_{0} + \theta_{-} \right)\right) - \frac{1}{z_{2} - 1}\biggl( \frac{1}{3} - \frac{1}{2}\left(\theta_{-} + \theta_{+}\right)\biggr)
- \frac{1}{z_{2} - z_{1}}\biggl( \frac{1}{3} - \frac{1}{2}\left(\theta_{1} + \theta_{-}\right) \biggr),
\end{equation}
\end{subequations}
and the separation constant is
\begin{equation}
\tilde{\lambda} = \frac{L^{2}\lambda}{\left(1-A^{2}L^{2}\right)(r_{+} - r_{0})(r_{-}-r_{1})},
\end{equation}
where  $\lbrace z_{1}, z_{2}, K_{1}, K_{2} \rbrace$ are the conformal moduli and the accessory parameters of a Heun-like equation, respectively. Notice that the asymptotic behavior of the solutions around $z = \infty$ is given by
\begin{equation}
f^{(1)}_{\infty}(z) = z^{-\kappa_{1}}\left(1 + \mathcal{O}\left(z^{-1}\right)\right), \qquad f^{(2)}_{\infty}(z) = z^{-\kappa_{2}}\left(1 + \mathcal{O}\left(z^{-1}\right)\right),
\end{equation}
where
\begin{equation}\label{eq:theta_infty}
\theta_{\infty} \equiv \kappa_{1} - \kappa_{2}
\end{equation}
and the regularity at $z = \infty$ implies the Fuchs condition
\begin{equation}\label{eq:fuchs}
\kappa_{1} + \kappa_{2} + \theta_{0} + \theta_{1} + \theta_{-} + \theta_{+} = 3.
\end{equation} 
\subsubsection{Boundary Conditions}
\label{sec:2.1.3}

The QNMs are solutions of the eigenvalue problem relative to \eqref{eq:radial_ode} obeying specific boundary conditions. Namely, the scattering region is defined between $r_{+} < r < \infty$, and we impose purely ingoing wave at the outer horizon and regularity at the spatial infinity, which implies the following asymptotic behavior
\begin{equation}\label{eq:boundaryforr}
R(r)\sim
\begin{cases}
(r-r_{+})^{-\theta_{+}/2},& r \rightarrow r_{+}, \\
\\
A r^{-2} + B r^{-1},& r \rightarrow \infty.
\end{cases}
\end{equation}
where $A$ and $B$ are constants. In addition, for the angular equation \eqref{eq:angular_ode}, we are interested in solutions that satisfy 
\begin{equation}\label{eq:boundaryfortheta}
S(w)\sim
\begin{cases}
w^{\rho_{-1}/2},& w \rightarrow 0 \,\left(\theta \rightarrow \pi\right), \\
\\
(w - w_{1})^{\rho_{1}/2},& w \rightarrow w_{1}\,\left(\theta \rightarrow 0\right),
\end{cases}
\end{equation}

i.e., they do not blow up in the domain $\theta \in \left[0,\pi\right]$. Equations \eqref{eq:heun_ang_conformal} and \eqref{eq:heun_radial}, along with the boundary conditions \eqref{eq:boundaryforr} and \eqref{eq:boundaryfortheta}, establish the eigenvalue problem for the radial and angular system (see \cite{Berti:2009kk,Konoplya:2011qq,Konoplya:2013rxa} for further discussions). In the following section, we will solve the angular eigenvalue problem by determining the separation constant $\lambda$; then, we will substitute it into the radial ODE \eqref{eq:heun_radial} to calculate the QNMs frequencies.

\section{The isomonodromic tau function of the accelerating KNAdS$_{4}$ black hole}
\label{sec:3}

The boundary value problem can be recast in terms of an initial value problem of an isomonodromic tau function. Namely, it turns out that the initial conditions for the Painlev\'{e} VI tau function (see \cite{Gamayun:2012ma} for the series representation of this transcendental function) solve the eigenvalue problem associated with the Heun equation as shown in \cite{CarneirodaCunha:2015qln}. Furthermore, the stability of several black holes under linear perturbations has been studied by solving numerically the set of transcendental equations, and asymptotic expansions for the QNMs frequencies have been obtained in the slowly rotating and the small black hole limit \cite{Novaes:2018fry,BarraganAmado:2018zpa,Amado:2021erf}. Other Painlev\'{e} tau functions have been introduced to study the confluent Heun equations that result after separating the equations of motion in the presence of asymptotically flat black hole solutions \cite{CarneirodaCunha:2015hzd,CarneirodaCunha:2019tia,Cavalcante:2021scq}. Nonetheless, the presence of a fifth regular singular point has been only recently addressed in the context of conformal mapping of polycircular arc domains \cite{CarneirodaCunha:2021jsu}. By using the definition of the isomonodromic tau function in terms of the Fredholm determinant \cite{Gavrylenko:2016zlf}, the authors implemented the tau function associated with Fuchsian systems of five regular singular points. To our knowledge, the same type of ODEs appears in the scattering off massive scalar and fermionic fields by Kerr-Newman-(A)dS black holes in four dimensions \cite{Kraniotis:2016maw,Kraniotis:2018zmh}.

In the following section, we will address the angular eigenvalue problem in the slowly rotating and accelerating black hole limit. Since the angular ODE  \eqref{eq:heun_ang_conformal} possesses four regular singular points, one can reformulate the boundary value problem using a $2 \times 2$ first-order linear system. Then, by demanding isomonodromy, the Schlesinger equations reduce to the Painlev\'{e} VI equation for an apparent singularity $\lambda$ --see \eqref{eq:PainleveVI}. In fact, the latter equation admits a Hamiltonian structure. Given initial conditions for the dynamical system \eqref{eq:initial_conditions_PVI}, one recovers the angular Heun equation and establish a map between its parameters $\lbrace w_{1}, Q_{1} \rbrace$ and the Painlev\'{e} VI tau function related to the monodromy data of Fuchsian systems with four singularities.

\subsection{The separation constant}
\label{sec:3a}
We consider a $2 \times 2$ first-order linear system with four regular singular points on the Riemann sphere:
\begin{equation}\label{eq:matrixODE}
\dfrac{d\Phi}{dz} = A(z)\Phi, \qquad z \in \mathds{C},
\end{equation}
where $\Phi(z)$ is the fundamental matrix of solutions defined in the complex plane 
\begin{equation}\label{eq:Phi}
\Phi(z) = \begin{pmatrix}
y_{1}(z) & w_{1}(z) \\
y_{2}(z) & w_{2}(z) \\
\end{pmatrix},
\end{equation}
and $A(z)$ is given by a partial fraction expansion of the form
\begin{subequations}
\begin{equation}\label{eq:Fuchsian_four}
\dfrac{d\Phi}{dz}= \left[\dfrac{A_{0}}{z} + \dfrac{A_{t}}{z - t} + \dfrac{A_{1}}{z - 1}\right]\Phi(z),
\end{equation}
with
\begin{equation}
A_{\infty}=-\left(A_{0} + A_{t} + A_{1}\right),
\end{equation}
\end{subequations}
where $A_{i} \, \left(i = 0, t, 1 \right)$ are $2 \times 2$ residue matrices that do not depend on $z$ and, as a consequence of the isomonodromy condition, satisfy the Schlesinger equations
\begin{equation}\label{eq:schlesinger}
		\frac{d A_{0}}{dt} = -\frac{\left[A_{0},A_{t}\right]}{t}, \qquad
		\frac{d A_{1}}{dt} = -\frac{\left[A_{1},A_{t}\right]}{t - 1}, \qquad
		\frac{d A_{t}}{dt} = \frac{\left[A_{0},A_{t}\right]}{t} + \frac{\left[A_{1},A_{t}\right]}{t - 1},
\end{equation}
where the last equation is equivalent to $A_{\infty} = \mathrm{const}$. Using the freedom of global conjugation, one may assume
\begin{equation}\label{eq:Ainfity}
A_{\infty}=
\begin{pmatrix}
\vartheta_{\infty} & 0 \\
0 & -\vartheta_{\infty}
\end{pmatrix},
\end{equation}
and it turns out that the entry $A_{12}(z)$ of $A(z)$ has only one zero in the complex plane and is given by
\begin{equation}\label{eq:non_diagonal_entry}
A_{12}(z) = \dfrac{k(z-\lambda)}{z(z-1)(z-t)},
\end{equation}
with $k \in \mathds{C}$. Surprisingly, the position of the zero of $A_{12}(z)$ with respect to the position of the pole at $z=t$ can be seen as a function $\lambda(t)$ which satisfies the Painlev\'{e} VI equation,
\begin{align}\label{eq:PainleveVI}
\lambda^{\prime\prime} = &\dfrac{1}{2}\left(\dfrac{1}{\lambda} + \dfrac{1}{\lambda-1} + \dfrac{1}{\lambda-t}\right)\left(\lambda^{\prime}\right)^{2}-\left(\dfrac{1}{t}+\dfrac{1}{t-1}+\dfrac{1}{\lambda-t}\right)\lambda^{\prime} + \nonumber \\
&+ \dfrac{\lambda(\lambda-1)(\lambda-t)}{2t^{2}(t-1)^{2}}\left[\left(\vartheta_{\infty}-1\right)^{2} -\dfrac{\vartheta^{2}_{0}t}{\lambda^{2}} + \dfrac{\vartheta^{2}_{1}(t-1)}{(\lambda-1)^{2}} - \dfrac{(\vartheta^{2}_{t} - 1)t(t-1)}{(\lambda-t)^{2}}\right],
\end{align} 
where $\vartheta_{0,t,1}$ are the eigenvalues of the residue matrices $A_{i}$. Moreover, the Fuchsian system \eqref{eq:Fuchsian_four} has an associated second-order ODE for $y_{1}$ of the form

\begin{align}\label{eq:deformed_heun}
y^{\prime\prime} + \biggl(\dfrac{1 - \vartheta_{0}}{z} +\dfrac{1 - \vartheta_{t}}{z-t} + \dfrac{1-\vartheta_{1}}{z-1}-\dfrac{1}{z-\lambda}\biggr)y^{\prime} + \biggl[ &\dfrac{\frac{1}{4}\left((\vartheta_{0} + \vartheta_{t} + \vartheta_{1} + 1)^2 - (\vartheta_{\infty} - 1)^2\right)}{z(z-1)} \nonumber \\
&-\dfrac{t(t-1)K}{z(z-1)(z-t)}+\dfrac{\lambda(\lambda-1)\mu}{z(z-1)(z-\lambda)}\biggr]y = 0,
\end{align}

where the subscript $1$ in $y_{1}$ has been omitted. One can further show that the indicial exponents at $z = \lambda$ are integers $\left( 0,2 \right)$, so the point corresponds to an apparent singularity of the deformed Heun equation, and $\mu$ is the canonical variable conjugated to $\lambda$, with the Hamiltonian of the dynamical system is given by 

\begin{equation}\label{eq:Kamiltonian}
\begin{split}
K(\mu,\lambda,t) = \frac{\lambda(\lambda-1)(\lambda-t)}{t(t-1)}\biggl[
\mu^2 - \left(\frac{\vartheta_{0}}{\lambda} + \frac{\vartheta_{t} - 1}{\lambda-t} + \frac{\vartheta_{1}}{\lambda-1}
\right)\mu +\frac{\frac{1}{4}\left((\vartheta_{0} + \vartheta_{t} + \vartheta_{1} + 1)^2 -(\vartheta_{\infty} - 1)^2\right)}{\lambda(\lambda-1)}\biggr],
\end{split}
\end{equation}
which corresponds to the Hamiltonian of the Painlev\'{e} VI equation. If we assume that $\left( \lambda(t),\mu(t) \right)$ are PVI solutions, we can impose the following initial conditions for the isomonodromic flow

\begin{equation}\label{eq:initial_lamda_mu}
t = t_0, \qquad \lambda(t_0) = t_0, \qquad \mu(t_0) = \dfrac{K_{0}}{1-\theta_{t}},
\end{equation}
with the identification
\begin{equation}
\vartheta_{0}= \theta_{0}, \qquad \vartheta_{t}= \theta_{t} - 1, \qquad \vartheta_{1}= \theta_{1}, \qquad \vartheta_{\infty}  = \theta_{\infty} + 1,
\end{equation}
such that the deformed Heun equation \eqref{eq:deformed_heun} reduces to a Heun equation

\begin{equation}\label{eq:heun_equation}
y^{\prime\prime} + \biggl(\dfrac{1 - \theta_{0}}{z} + \dfrac{1 - \theta_{t}}{z - t_{0}} + \dfrac{1 -\theta_{1}}{z - 1}\biggr)y^{\prime} + \biggl[ \dfrac{\frac{1}{4}\left((\theta_{0} + \theta_{t} + \theta_{1})^2 - \theta_{\infty}^2\right)}{z(z - 1)}
-\dfrac{t_{0}(t_{0} - 1)K_{0}}{z(z - 1)(z - t_{0})}\biggr]y = 0,
\end{equation}
which coincides with \eqref{eq:heun_ang_conformal}, and gives the relation between the Heun equation and the angular equation parameters
\begin{table}[H]
\centering
\begin{tabular}{c c c c c c c}
\midrule 
Eq.\eqref{eq:heun_equation} & $t_{0}$ & $\theta_{0}$ & $\theta_{t}$ & $\theta_{1}$ & $\theta_{\infty}$ & $K_{0}$ \\  
\midrule 
Eq.\eqref{eq:heun_ang_conformal} & $w_{1}$ & $\rho_{-1}$ & $\rho_{1}$ & $\rho_{u_{0}}$ & $\rho_{u_{1}}$ & $Q_{1}$ \\
\midrule
\end{tabular}
\caption{Map from the Heun equation to the angular equation parameters.}
\label{table:1}
\end{table}
By means of an isomonodromic tau function in the Jimbo, Miwa and Ueno sense, defined as
\begin{equation}
\frac{d}{d t}\log \tau(\rho;t) = \frac{1}{t}\mathrm{Tr}\,A_{0}A_{t} + \frac{1}{t - 1}\mathrm{Tr}\,A_{t}A_{1}
\end{equation}
where $\vartheta = \lbrace \vartheta_{0},\vartheta_{t},\vartheta_{1},\vartheta_{\infty} \rbrace$ are the local monodromy exponents, and $\rho = \lbrace \vartheta; \sigma,\eta \rbrace$ are the monodromy data associated with the $2 \times 2$ Fuchsian system with four singularities, one can rewrite the initial conditions \eqref{eq:initial_lamda_mu} in terms of the PVI tau function as follows

\begin{subequations}\label{eq:initial_conditions_PVI}
\begin{equation}\label{eq:logderivative}
\frac{d}{d t}\log\tau_{\rm VI}\left(\rho;t\right)\biggr|_{t=w_{1}} = \frac{\left(\rho_{1}-1\right)\rho_{u_{0}}}{2(w_{1}-1)}+\frac{\left(\rho_{1}-1\right)\rho_{-1}}{2w_{1}} + Q_{1},
\end{equation}
\begin{equation}\label{eq:toda}
\tau_{\rm VI}\left(\rho^{+};w_{1}\right) = 0,
\end{equation}
\end{subequations}
where the monodromy arguments are given as follows:
\begin{equation}\label{eq:monodromy_PVI}
\rho = \lbrace \vartheta;\sigma,\eta \rbrace = \lbrace \rho_{-1}, \rho_{1} - 1, \rho_{u_{0}}, \rho_{u_{1}} + 1; \sigma - 1; \eta \rbrace,
\end{equation}
Furthermore, the monodromy parameters in \eqref{eq:toda} are related by shifts on the monodromies \eqref{eq:monodromy_PVI}
\begin{equation}\label{eq:monodromy_toda}
\rho^{+} = \lbrace \vartheta^{+};\sigma^{+},\eta^{+} \rbrace = \lbrace \rho_{-1}, \rho_{1}, \rho_{u_{0}}, \rho_{u_{1}}; \sigma; \eta \rbrace.
\end{equation}

In the case of slowly accelerating and rotating black holes, we can use the series expansion of the tau function around zero \cite{Gamayun:2012ma,Gamayun:2013auu}. Therefore, the second condition \eqref{eq:toda} can be inverted for $t = w_{1}$ sufficiently close to zero, and its expansion can be computed recursively. Then, the logarithmic derivative of the PVI tau function in \eqref{eq:logderivative} gives an analytic expansion of the accessory parameter in terms of the monodromy data and the conformal modulus 

\begin{equation}\label{eq:fourt0K0}
\begin{gathered}
4 w_{1} Q_{1} = \left(\sigma - 1\right)^{2} - \left(\rho_{-1} + \rho_{1} - 1\right)^{2} + \biggl[2(\rho_{u_{0}} - 1)(\rho_{1} - 1)
+ \frac{\left((\sigma - 1)^2 + \rho_{1}^2 - \rho_{-1}^2 - 1\right)\left((\sigma - 1)^2 + \rho_{u_{0}}^2 -\rho_{u_{1}}^2 - 1\right)}{2\sigma(\sigma - 2)}\biggr]w_{1}\\
+\biggl[\frac{13}{32}\sigma(\sigma - 2) + 2\left(\rho_{u_{0}} - 1\right)\left(\rho_{1} - 1\right) - \frac{1}{32}\left(5 + 14\left(\rho^{2}_{-1} + \rho^{2}_{u_{1}}\right) - 18\left(\rho^{2}_{1} + \rho^{2}_{u_{0}}\right)\right) + \frac{\left(\rho^{2}_{-1} - \rho^{2}_{1}\right)^{2}\left(\rho^{2}_{u_{0}} - \rho^{2}_{u_{1}}\right)^{2}}{64}\left(\frac{1}{\sigma^{3}} - \frac{1}{(\sigma-2)^{3}}\right)\\
-\frac{\left(\left(\rho^{2}_{-1} - \rho^{2}_{1}\right)\left(\rho^{2}_{u_{0}} - \rho^{2}_{u_{1}}\right) + 8\right)^{2} - 2\left(\rho^{2}_{-1} + \rho^{2}_{1}\right)\left(\rho^{2}_{u_{0}} - \rho^{2}_{u_{1}}\right)^{2} - 2\left(\rho^{2}_{-1} - \rho^{2}_{1}\right)^{2}\left(\rho^{2}_{u_{0}} + \rho^{2}_{u_{1}}\right) - 64}{32\sigma(\sigma - 2)}\\
+ \frac{\left((\rho_{-1} - 1)^{2} - \rho^{2}_{1}\right)\left((\rho_{-1} + 1)^{2} - \rho^{2}_{1}\right)\left((\rho_{u_{0}} - 1)^{2} - \rho^{2}_{u_{1}}\right)\left((\rho_{u_{0}} + 1)^{2} - \rho^{2}_{u_{1}}\right)}{32(\sigma + 1)(\sigma - 3)}
\biggr]w^{2}_{1} +\mathcal{O}(w_{1}^{3}).
\end{gathered}
\end{equation}

Hence, we substitute \eqref{eq:new_accessory} into \eqref{eq:fourt0K0} to obtain an asymptotic expansion for the separation constant. Nonetheless, we still need to impose the angular quantization condition on $\sigma$, which can be read from the boundary conditions \eqref{eq:boundaryfortheta}. By demanding that the separation constant in the case of non-accelerating Schwarzschild-AdS$_{4}$, i.e., for $a \rightarrow 0$, $e \rightarrow 0$, and $A \rightarrow 0$, is 
\begin{equation}\label{eq:lambda_conformal}
\lambda_{\ell} = \ell\left(\ell + 1\right) + \frac{1}{3},
\end{equation}
we can obtain the following conditions
\begin{equation}\label{eq:sigma_ell}
\sigma = 2(\ell+1), \qquad \sigma = -2\ell, \qquad \ell \geq \vert m \vert,
\end{equation}
where $\ell$ is the angular momentum quantum number. 
Choosing the first angular quantization condition and taking the slowly accelerating and rotating limit, the separation constant of the accelerating Kerr-Newman-AdS$_{4}$ black hole yields\footnote{Notice that by replacing $\bar{A} = L A$, $\bar{a} = a/L$, $\bar{e} = e/L$, and $\bar{r}_{+} = r_{+}/L$, the AdS radius can be removed from the expression \eqref{eq:lambda_knads}.}

\begin{equation}\label{eq:lambda_knads}
\begin{split}
\lambda_{\omega,\ell,m} &= \ell\left(\ell + 1\right) + \frac{1}{3} - 2m\left(\omega + \frac{e q}{r_{+}}\right)a  + 2m \frac{(\omega r_{+} + e q)(r_{+}^{4} + L^{2}r_{+}^{2} + e^{2}L^{2})}{L^{2}r_{+}^{2}}a\,A \\
&+\frac{6L^{2}\left(\ell(\ell+1) + (m+1)(m-1)\right)(\omega r_{+} + e q)^{2} - r_{+}^{2}\left(6m^2\left(3\ell^{2}+3\ell-2\right)+6\ell^{4}+12\ell^{3}+5\ell^{2}-\ell-3\right)}{3L^{2}r_{+}^{2}(2\ell-1)(2\ell+3)}a^{2} \\
&+ \biggl[ \frac{(1-\ell(\ell+1)(3\ell^{2}+3\ell-1)-m^{2}(15\ell^{2}+15\ell-11))}{2(2\ell-1)(2\ell+3)}r_{+}^{2}\left(1+\frac{2r_{+}^{2}}{L^{2}}+\frac{r_{+}^{4}}{L^{4}} + \frac{2e^{2}}{L^{2}} + \frac{e^{4}}{r_{+}^{4}}\right) \\
&- \frac{\left(\ell(\ell+1)(3\ell^{2}+3\ell-2) + 3m^{2}(9\ell^{2}+9\ell-7)\right)}{3(2\ell-1)(2\ell+3)}e^{2}\biggr]A^{2} + \mathcal{O}\left(a^{3},a^{2}A,a A^{2},A^{3}\right)
\end{split}
\end{equation}

Interestingly, the angular eigenvalues for different black solutions can be recovered by setting the rotation parameter, the charge or the acceleration to zero. 
It reduces to the separation constant of an accelerating Reissner-Nordstr\"{o}m-AdS$_{4}$ black hole for $a \rightarrow 0$,

\begin{equation}\label{eq:lambda_accrnads}
\begin{split}
\lambda_{\ell,m} &= \ell\left(\ell + 1\right) + \frac{1}{3} + \biggl[ \frac{(1-\ell(\ell+1)(3\ell^{2}+3\ell-1)-m^{2}(15\ell^{2}+15\ell-11))}{2(2\ell-1)(2\ell+3)}r_{+}^{2}\left(1+\frac{2r_{+}^{2}}{L^{2}}+\frac{r_{+}^{4}}{L^{4}} + \frac{2e^{2}}{L^{2}} + \frac{e^{4}}{r_{+}^{4}}\right) \\
&- \frac{\left(\ell(\ell+1)(3\ell^{2}+3\ell-2) + 3m^{2}(9\ell^{2}+9\ell-7)\right)}{3(2\ell-1)(2\ell+3)}e^{2}\biggr]A^{2} + \mathcal{O}(A^{3}), 
\end{split}
\end{equation}
which reproduces the numerical results presented in \cite{Fontana:2022whx}. Moreover, it also gives the angular eigenvalues of the accelerating Schwarzschild-AdS$_{4}$ when additionally $e \rightarrow 0$. The initial conditions \eqref{eq:initial_conditions_PVI} are valid for generic $w_{1}$, and the accuracy of the computation of the angular eigenvalues depends on the implementation of the PVI tau function. However, due to the combinatorial nature of the Nekrasov functions, the coefficients in the series expansion of the PVI tau function are cumbersome. Therefore, we have only written the separation constant up to the second order in $A$. 

\subsection{Quasinormal modes}
\label{sec:3b}

Once the separation constant has been computed, the eigenvalue problem for the radial system can be fully solved and we can regard the QNM frequencies as solutions to \eqref{eq:heun_radial} that obey the boundary conditions \eqref{eq:boundaryforr}.
Since we are interested in applying the method of isomonodromic deformations, we will consider a $2 \times 2$ first-order linear system with five regular singular points, as described in Appendix \ref{appendix_B}. It turns out that the isomonodromy approach requires the addition of two apparent singularities \eqref{eq:A_12} --while the in the case of Fuchsian sytems with four singularities, we only need to introduce one \eqref{eq:non_diagonal_entry}-- to match the presence of two isomonodromic deformation parameters $\lbrace t_{1}, t_{2} \rbrace$. Nevertheless, the isomonodromic deformation equations can be written as a Hamiltonian system \eqref{eq:Hamilton_eqns} which is completely integrable. 

\begin{figure}[H]
\centering
\includegraphics[width=0.425\linewidth]{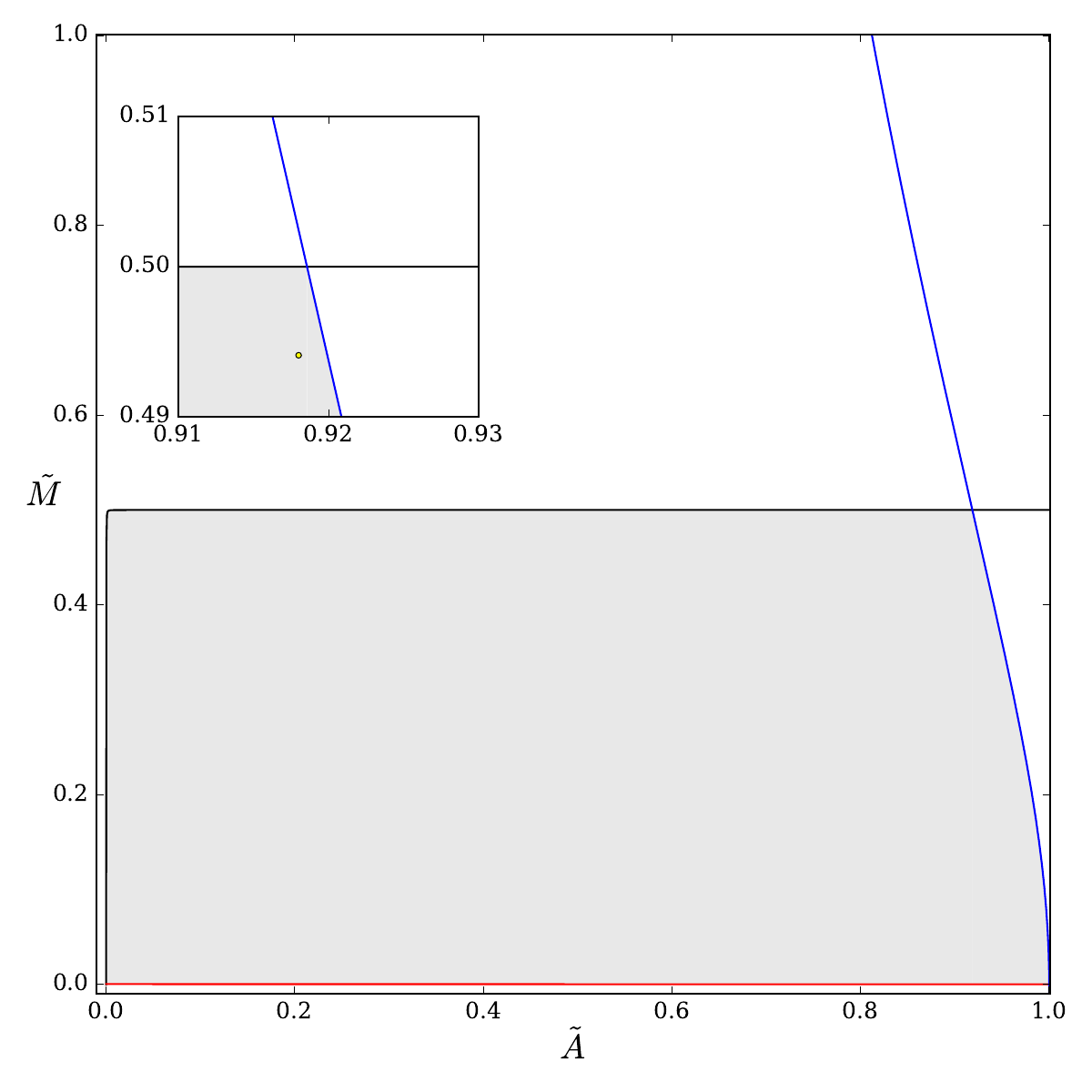}
\caption{The $\tilde{A}-\tilde{M}$ parameter space of slowly accelerating Kerr-Newman-AdS$_{4}$ black holes for $a = e = 10^{-4}$ and $L=1$. The inset plot at the left corner zooms in on the intersection point between the black and blue lines, the region of slowly accelerating black holes where we have explored the computation of QNMs.}
\label{fig:space_slowly_acc_bh}
\end{figure}

One can see that the by imposing the following changes on the local monodromy exponents and initial conditions for the Hamiltonian system \eqref{eq:Hamilton_eqns}, the deformed Heun equation \eqref{eq:deformed_five} recovers the Heun-like equation \eqref{eq:heun-like_equation}. Namely, we have 
\begin{equation}\label{eq:initial_conditions_garnier}
\begin{gathered}
\vartheta_{0} = \theta_{0}, \quad \vartheta_{1} = \theta_{1}-1, \quad \vartheta_{2} = \theta_{2}-1, \quad \vartheta_{3} = \theta_{3}, \quad \vartheta_{\infty} = \theta_{\infty}+1, \\
\lambda_{1}(t_{1}=z_{1},t_{2}=z_{2}) = z_{1}, \quad \mu_{1}(t_1=z_{1},t_2=z_{2}) = \frac{K_{1}}{1-\theta_{1}},\\
\lambda_{2}(t_{1}=z_{1},t_{2}=z_{2}) = z_{2}, \quad \mu_{2}(t_1=z_{1},t_2=z_{2}) = \frac{K_{2}}{1-\theta_{2}}.
\end{gathered}
\end{equation}
which coincides with \eqref{eq:heun_radial} under the radial dictionary  
\begin{table}[H]
\centering
\begin{tabular}{c c c c c c c c c c} 
\midrule 
Eq.\eqref{eq:heun-like_equation} & $z_{1}$ & $z_{2}$ & $\theta_{0}$ & $\theta_{1}$ & $\theta_{2}$ & $\theta_{3}$ & $\theta_{\infty}$ & $K_{1}$ & $K_{2}$\\ 
\midrule 
Eq.\eqref{eq:heun_radial} & $z_{1}$ & $z_{2}$ & $\theta_{0}$ & $\theta_{1}$ & $\theta_{-}$ & $\theta_{+}$& $1$ & $K_{1}$ & $K_{2}$ \\ 
\midrule
\end{tabular}
\caption{Map from the Heun-like equation to the radial equation parameters.}
\label{table:2}
\end{table}
These conditions can be expressed in terms of the isomonodromic tau function yielding
\begin{subequations}\label{eq:transcendental}
\begin{equation}\label{eq:dlog_tau1}
\frac{\partial}{\partial t_{1}}\log \tau_{\mathrm{JMU}}\left(\rho;z_{1},z_{2} \right) = K_{1} 
+ \frac{\theta_{0}(\theta_{1} - 1)}{2 z_{1}} + \frac{(\theta_{1} - 1)(\theta_{-} - 1)}{2(z_{1} - z_{2})} + \frac{(\theta_{1} - 1)\theta_{+}}{2(z_{1} - 1)},
\end{equation}
\begin{equation}\label{eq:dlog_tau2}
\frac{\partial}{\partial t_{2}}\log \tau_{\mathrm{JMU}}\left(\rho;z_{1},z_{2} \right) = K_{2} 
+ \frac{\theta_{0}(\theta_{-} - 1)}{2 z_{2}} + \frac{(\theta_{-} - 1)(\theta_{1} - 1)}{2(z_{2} - z_{1})} + \frac{(\theta_{-} - 1)\theta_{+}}{2(z_{2} - 1)},
\end{equation}
\begin{equation}\label{eq:zero_tau1}
\tau_{\rm JMU}\left(\rho^{+}_{1};z_{1},z_{2}\right) = 0,
\end{equation}
\begin{equation}\label{eq:zero_tau2}
\tau_{\rm JMU}\left(\rho^{+}_{2};z_{1},z_{2}\right) = 0,
\end{equation}
\end{subequations}
supplemented by the monodromy data $\rho = \lbrace \vartheta; \sigma_{k}; \eta_{k} \rbrace, \left(k = 1,2\right)$ as defined in \eqref{eq:monodromy_data}, with the local monodromy exponents, the conformal moduli, and the accessory parameters replaced according to Table \ref{table:2}. 
The local monodromy exponents obey $\theta^{+}_{\nu} - \theta^{-}_{\nu} \notin \mathbb{Z}$, such that the monodromy matrices can be written as
\begin{equation}\label{eq:M_nu}
M_{\nu} = g_{\nu}^{-1}\begin{pmatrix}
e^{2\pi i \theta_{\nu}} & 0 \\
0 & e^{-2\pi i \theta_{\nu}} \\
\end{pmatrix}g_{\nu}.
\end{equation}
Interestingly, at the point $z = \infty$, the Frobenius exponents differ by an integer $\theta_{\infty} = 1$, which can introduce logarithmic singularities in the solutions. However, one can circumvent this issue if the diagonal parts corresponding to the degenerate eigenvalues are replaced by the appropriate Jordan blocks, generating a simplification in the Fredholm determinant representation (see \cite{Gavrylenko:2016zlf} for a discussion on the $4$-point tau function). The latter consideration will affect the trinion $\mathcal{T}^{[3]}$ in the pants decomposition $\eqref{eq:trinions}$, since it contains the local monodromy at infinity. For instance, one may set the monodromy matrices at $z = 1$ and $z = \infty$ to have a diagonal and an upper triangular form
\begin{equation}\label{eq:red_trinion}
J_{1} = \begin{pmatrix}
e^{2\pi i \theta_{+}} & 0 \\
0 & e^{-2\pi i \theta_{+}} \\
\end{pmatrix},
\qquad
J_{\infty} = \begin{pmatrix}
1 & 1 \\
0 & 1 \\
\end{pmatrix},
\end{equation}
such that a general solution can be written as $\left( R^{(\rm reg)} \, R^{(\rm irr)}\right) = R^{(\infty)}g_{\infty}$, where $M_{\infty} = g_{\infty}^{-1}J_{\infty}g_{\infty}$. Then, we use the basis $\Tr \left( M_{1}M_{\infty}\right) = -2\cos\left(2\pi\sigma_{2}\right)$, and express $\mathcal{M}_{1\infty} = g_{1} g_{\infty}^{-1}$ as the connection matrix between the solution around the outer horizon and the spatial infinity. In other words, $\sigma_{2}$ encloses the singularities $z = 1 (r = r_{+})$ and $z = \infty (r = \infty)$, so that it can be determined using the radial boundary conditions \eqref{eq:boundaryforr} as shown in \cite{BarraganAmado:2018zpa,Amado:2021erf}. Then, the regularity condition gives the composite monodromy around these points
\begin{equation}\label{eq:rad_quantization}
\sigma_{2} = \theta_{+} + 2n, \qquad n \in \mathds{Z}.
\end{equation}
Thus, using an asymptotic expansion for the isomonodromic tau function \eqref{eq:tau_jmu} into the set of transcendental equations \eqref{eq:transcendental} gives a procedure to compute the QNMs. For $n = 5$, the tau function \eqref{eq:tau_five} will only depend on  $\lbrace \omega, \sigma_{1}, \eta_{1}, \eta_{2} \rbrace$, as the local monodromies \eqref{eq:thetas} are functions of the frequency. In order to find the eigenfrequencies and the separation constant, we will numerically solve the radial and angular systems of equations \eqref{eq:transcendental} and \eqref{eq:initial_conditions_PVI} respectively. It is worth mentioning that for black holes that exist inside the parameter space of the slowly accelerating limit such as the black hole configuration represented by the yellow point in Figure \ref{fig:space_slowly_acc_bh}, the radial conformal moduli $z_{1}, z_{2}$ are complex numbers as shown in Table \ref{table:3}. However, we can still implement the series expansion of the operators $\mathsf{a}^{[k]},\mathsf{b}^{[k]},\mathsf{c}^{[k]},$ and $\mathsf{d}^{[k]}$ in the asymptotic expansion of the isomonodromic tau function \eqref{eq:tau_five}; namely, the tau function \eqref{eq:tau_five} is truncated up to order $Q = 56$. On the other hand, in the angular system, the chosen black hole parameters will produce a $w_{1}$ close to one. Therefore, we need to perform the series expansion of the PVI tau function around $t = 1$, where the relevant composite monodromy $\left(\sigma_{t1} = \bar{\sigma}\right)$ encloses the points $w = w_{1}$ and $w = 1$, in contrast with the expansion around $t = 0$ where the appropriate composite monodromy $\left( \sigma_{0t} = \sigma \right)$ encircles the points $w = 0$ and $w = w_{1}$. 

\begin{figure}[!htb]
\centering
\includegraphics[width=0.85\linewidth]{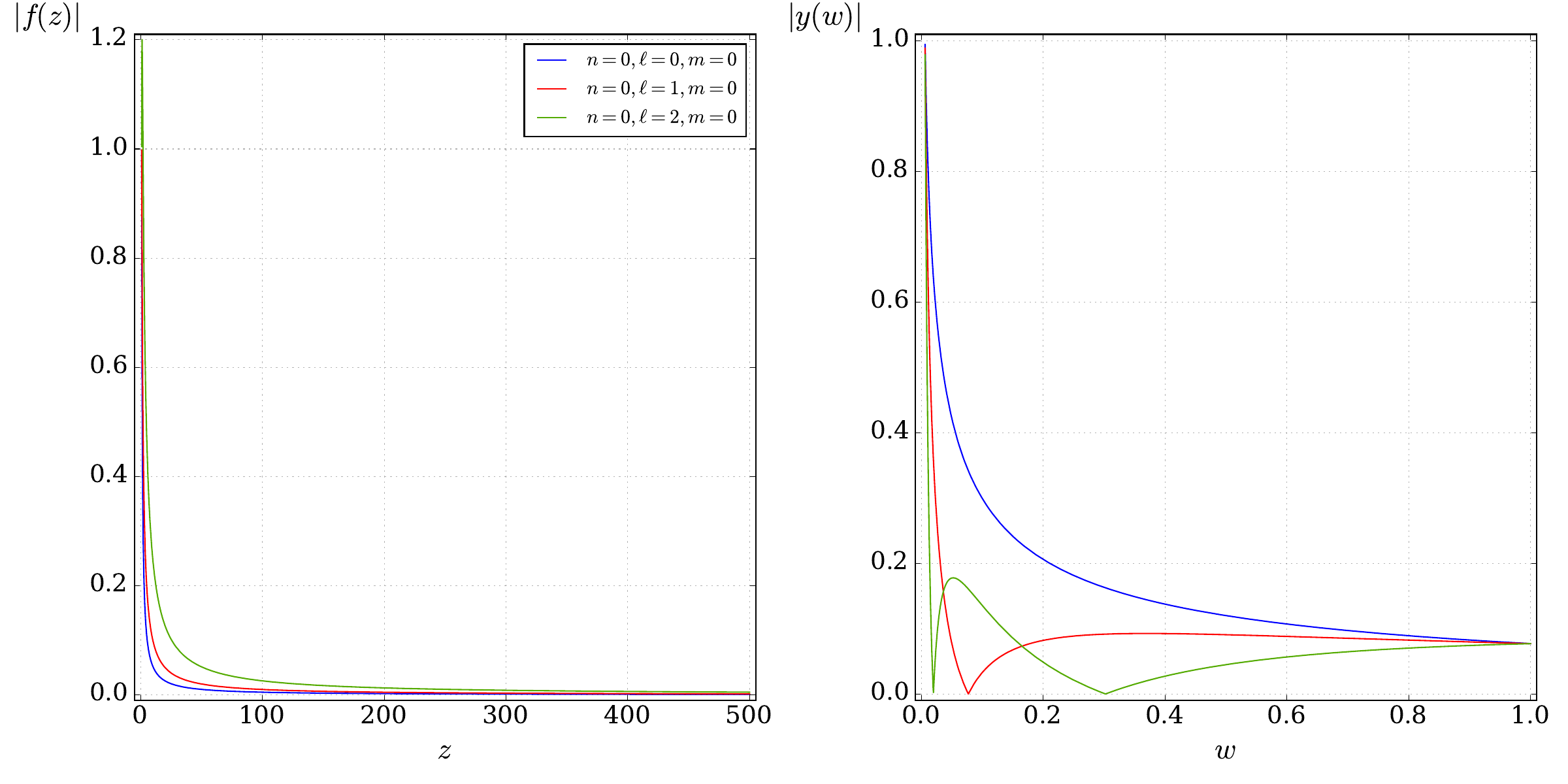}
\caption{Numerical solutions using the Direct Integration method for the radial $\vert f(z) \vert$ (left panel) and the angular ODE $\vert y(w) \vert$ (right panel) with parameters defined in Table \ref{table:3} and different $\ell$.}
\label{fig:sols}
\end{figure}

\begin{table}[H]
\centering
\begin{tabular}{c c c c c}
\midrule
$r_{+}$	& $z_{1}$                   & $z_{2}$                   & $z_{1}/z_{2}$             & $w_{1}$    \\ \midrule
$0.65$  & $0.03071661 + 0.2459466i$ & $0.8341494 + 0.02071331i$ & $0.04411822 + 0.2937516i$ & $0.9940647$ \\ \midrule
\end{tabular}
\caption{Conformal moduli for the following black hole parameters: $a = e = 10^{-4}$, $A = 0.918$, and $L = 1$.}
\label{table:3}
\end{table}
In Figure \ref{fig:sols}, we solve the radial and angular ODEs using the direct integration method to verify the solutions for the QNMs frequencies and the separation constant obtained via the isomodromic tau function. The idea is to construct two Frobenius solutions near the horizon and spatial infinity, which satisfy continuity conditions at the matching point. As a result, correct eigenvalues $\lbrace \omega_{n,\ell,m}, \lambda_{\ell,m} \rbrace$ will produce a set of solutions $\lbrace f_{n,\ell,m}(z), y_{\ell,m}(w) \rbrace$, which have the correct asymptotic behavior and are continuous everywhere, as well as their first derivatives \cite{Molina:2010fb}. Despite the large value of $\mathrm{Re}(z_{2})$, the radial and angular solutions are continuous and smooth at the matching point, serving as validation of the isomonodromic deformation method. 

In Table \ref{table:4}, we present the numerical solutions of the initial conditions of the radial tau function \eqref{eq:transcendental} for different angular momentum quantum numbers and radius of the horizon of the black hole, keeping fixed the rotation, charge and acceleration parameters. It can be observed that for neutral scalar fields with $\ell = \lbrace 0,1,2 \rbrace$ and $m = 0$, the $n = 0$ modes are unstable in the sense that $\mathrm{Im}\,\omega_{0,\ell,0} > 0$, for $0.65 \leq r_{+} \leq 0.653$ and $A = 0.918$. Interestingly, the instability of the modes with $\ell = m = 0$ is not of superradiant nature since the superradiance condition, which involves the frequency, the angular momentum, and the charge of the scalar field, as well as the angular velocity of the event horizon and the electrostatic potential between the black hole and the spatial infinity, is satisfied \cite{Uchikata:2009zz,Uchikata:2011zz,Li:2012rx}. Therefore, a plausible explanation is that the black hole's acceleration can generate it. However, a better understanding of these modes is required to elucidate the origin of the instability. On the other hand, the set of initial conditions \eqref{eq:initial_conditions_PVI} provides the solution to $\lambda_{\ell,m}$ and $\bar{\sigma}$ as shown in Table \ref{table:5}. The complete numerical solution of the boundary value problem can be read off from Table \ref{table:4} and Table \ref{table:5}.

A thorough numerical analysis of the QNMs and the presence of superradiance is deserved, but is beyond the scope of this work, since we have merely attempted to describe the solution of the radial eigenvalue problem in terms of the isomonodromic tau function of Fuchsian systems with five regular singular points.

\label{appendix_D}
\begin{table}[!htb]
\centering
\scalebox{0.75}{
\begin{tabular}{c c c c c c c c}
\midrule
$\left( \ell, m \right)$ & $r_{+}$	& $z_{1}$                   & $z_{2}$                   & $\omega_{0,\ell,m}$           & $\sigma_{1}$ & $\eta_{1}$ & $\eta_{2}$  \\ \midrule
$\left(0,0\right)$       &$0.65$    & $0.03071661 + 0.2459466i$ & $0.8341494 + 0.02071331i$ & $-0.006351229 + 0.009373394i$ & $0.8726076 - 0.6982149i$ & $0.5730091 - 0.1134686i$ & $0.988138 + 0.00543559i$ \\
                         & $0.651$  & $0.02443598 + 0.2197154i$ & $0.8338561 + 0.01847795i$ & $-0.006121681 + 0.01004876i$  & $0.9047372 - 0.6995898i$ & $0.6234461 - 0.2225214i$ & $0.9883053 + 0.006679701i$ \\
                         & $0.652$  & $0.01813624 + 0.1895878i$ & $0.8335631 + 0.01592158i$ & $-0.005729512 + 0.01097167i$  & $0.9354836 - 0.70376i$   & $0.6356759 - 0.3693602i$ & $0.98863 + 0.008168491i$ \\
                         & $0.653$  & $0.01181725 + 0.1532803i$ & $0.8332706 + 0.01285411i$ & $-0.005036509 + 0.01244441i$  & $0.9637807 - 0.7211214i$ & $0.5224173 - 0.5601483i$ & $0.9892118 + 0.01005387i$ \\
						 &		    &			    			&                			&			                    &                          		    &                          &							\\ 
$\left(1,0\right)$       &$0.65$    & $0.03071661 + 0.2459466i$ & $0.8341494 + 0.02071331i$ & $-0.001243424 + 0.004198133i$ & $0.7729309 - 0.4701989i$ & $0.224334 + 0.522833i$ & $0.9940569 + 0.003826012i$ \\
                         & $0.651$  & $0.02443598 + 0.2197154i$ & $0.8338561 + 0.01847795i$ & $-0.001231874 + 0.004006211i$ & $0.8102078 - 0.4395744i$ & $0.2532531 + 0.6579928i$ & $0.9941022 + 0.004180008i$ \\
                         & $0.652$  & $0.01813624 + 0.1895878i$ & $0.8335631 + 0.01592158i$ & $-0.001228505 + 0.003793199i$ & $0.8506864 - 0.3993997i$ & $0.260636 + 0.8675718i$ & $0.9940839 + 0.004673116i$ \\
                         & $0.653$  & $0.01181725 + 0.1532803i$ & $0.8332706 + 0.01285411i$ & $-0.00128741 + 0.003718453i$  & $0.8928007 - 0.3494106i$ & $0.2100622 + 1.177701i$ & $0.9937097 + 0.005637174i$ \\
						 &		    &			    			&                			&			                    &                          		    &                          &							\\
$\left(2,0\right)$       &$0.65$    & $0.03071661 + 0.2459466i$ & $0.8341494 + 0.02071331i$ & $-0.0008263643 + 0.002847768i$ & $0.7865888 - 0.4520936i$ & $0.2121842 + 0.6007756i$ & $0.995887 + 0.002784429i$ \\
                         & $0.651$  & $0.02443598 + 0.2197154i$ & $0.8338561 + 0.01847795i$ & $-0.0008180703 + 0.002678499i$ & $0.8227354 - 0.4202813i$ & $0.226272 + 0.755103i$ & $0.9959489 + 0.002993748i$ \\
                         & $0.652$  & $0.01813624 + 0.1895878i$ & $0.8335631 + 0.01592158i$ & $-0.0008087563 + 0.002470681i$ & $0.8623027 - 0.3780972i$ & $0.2014858 + 0.9940133i$ & $0.9959926 + 0.003272807i$ \\
                         & $0.653$  & $0.01181725 + 0.1532803i$ & $0.8332706 + 0.01285411i$ & $-0.0008063592 + 0.002248942i$ & $0.9050444 - 0.3201995i$ & $0.0504816 + 1.369524i$ & $0.9959082 + 0.003773917i$ \\
						 &		    &			    			&                			&			                    &                          		    &                          &							\\						 
\midrule
\end{tabular}
}
\caption{Quasi-normal modes for neutral scalar fields with $n = 0, m = 0$ for different values of $r_{+}$ and angular momentum quantum numbers $\ell = \lbrace 0,1,2 \rbrace$, while keeping fixed $a = e = 10^{-4}$, $A=0.918$, and $L=1$.}
\label{table:4}
\end{table}
\begin{table}[!htb]
\centering
\scalebox{0.75}{
\begin{tabular}{c c c c c} 
\midrule
		$\left(\ell, m \right)$ & $r_{+}$ & $w_{1}$     & $\bar{\sigma}$ & $\lambda_{\ell,m}$ \\ 
\midrule
		$\left( 0, 0 \right)$   & $0.65$  & $0.9940647$ & $1 - 0.6095955i$ & $0.01896789$ \\
							    & $0.651$ & $0.9957813$ & $1 - 0.5889038i$ & $0.006695023$  \\
							    & $0.652$ & $0.9974996$ & $1 - 0.5599641i$ & $-0.009849679$   \\
							    & $0.653$ & $0.9992196$ & $1 - 0.5052231i$ & $-0.0389958$    \\
							    &		  &			    &                  &			     \\
		$\left( 1, 0 \right)$   & $0.65$  & $0.9940647$ & $1 - 1.301246i$  & $0.6739338$ \\
							    & $0.651$ & $0.9957813$ & $1 - 1.25159i$   & $0.6126858$ \\
							    & $0.652$ & $0.9974996$ & $1 - 1.182966i$  & $0.5310437$ \\
							    & $0.653$ & $0.9992196$ & $1 - 1.055796i$  & $0.3902286$ \\
							    &		  &			    &                  &		     \\
		$\left( 2, 0 \right)$   & $0.65$  & $0.9940647$ & $1 - 2.054172i$  & $1.925941$   \\
							    & $0.651$ & $0.9957813$ & $1 - 1.97173i$   & $1.765992$  \\
							    & $0.652$ & $0.9983594$ & $1 - 1.858023i$  & $1.553623$  \\
							    & $0.653$ & $0.9992196$ & $1 - 1.64821i$   & $1.190238$  \\
\midrule							  
\end{tabular}
}
\caption{Angular eigenvalues $\lambda$ of the accelerating KNAdS$_{4}$ black holes associated with the eigenfrequencies in Table \ref{table:4}.}
\label{table:5}
\end{table}

\section{Conclusions}
\label{sec:4}

The conformally coupled Klein-Gordon equation for a massless scalar field in \eqref{eq:conformal_metric} separates into two second-order ODEs for the radial and angular functions \eqref{eq:ODEs}. The resulting angular ODE can be transformed into a Heun equation \eqref{eq:heun_ang_conformal} as the singular point at $w = w_{2}$ can be removed from it by choosing $\beta^{-}_{\infty}$ in the s-homotopic transformation \cite{Suzuki:1998vy}. Therefore, the angular eigenvalue problem can be recast in terms of a set of initial conditions for the Painlev\'{e} VI tau function \eqref{eq:initial_conditions_PVI}. By expanding the accessory parameter equation \eqref{eq:fourt0K0}, we have derived an asymptotic expansion for the separation constant in the slowly accelerating and rotating limit \eqref{eq:lambda_knads}. This expression reproduces the results in the literature for the separation constant in the case of slowly accelerating Reissner-Nordstr\"{o}m-AdS black holes \cite{Fontana:2022whx}.

The radial ODE \eqref{eq:radial_ode} can be more intricate because we need to solve the second-order ODE with five regular singular points. In order to find the QNMs, we introduce the isomonodromic deformation of the $2 \times 2$ Fuchsian system with five singularities, which leads to a deformed Heun-like equation that possesses two extra apparent singularities \eqref{eq:deformed_five}. Then, by coalescing those apparent singularities with the two unfixed points, one can determine the initial conditions of the Hamiltonian system \eqref{eq:initial_conditions}, such that the deformed equation recovers the radial Heun-like equation \eqref{eq:heun_radial} under the appropriate radial dictionary --see Table \ref{table:2}. In turn, the initial conditions are written directly in terms of the isomonodromic tau function \eqref{eq:transcendental}, which provides the solution of the radial eigenvalue problem given the monodromy data.

In terms of the numerical results, we are interested in exploring the slowly accelerating KN-AdS$_{4}$ black hole limit. For the radial system \eqref{eq:transcendental}, we implement the asymptotic expansion of the Fredholm determinant of the isomonodromic tau function for $t_{1} \ll t_{2} \ll 1$. Nevertheless, for angular one, the conformal modulus $w_{1}$ is close to one, which makes it impossible to use the accessory parameter expansion derived in \eqref{eq:fourt0K0}. Therefore, we need to solve \eqref{eq:initial_conditions_PVI} using the series expansion of the PVI tau function around $t = 1$. As a result, the boundary value problem is recast in a set of six transcendental equations for $\lbrace \omega, \sigma_{1}, \eta_{1}, \eta_{2}, \lambda, \bar{\sigma} \rbrace$. Bearing this in mind, we have chosen black hole configurations close to one of the corners in the parameter space --as described by the yellow point in Figure \ref{fig:space_slowly_acc_bh}-- where $\mathrm{Re}(z_{1})$, $\mathrm{Re}(z_{2})$, and the ratio $\mathrm{Re}\left(z_{1}/z_{2}\right)$ are numerically tractable for the computation of the tau function.

Interestingly, we have found that the $(n=0)$ quasi-normal modes of neutral scalar fields for different angular momentum quantum numbers $\ell$ and horizon radius $r_{+}$, with $m = 0$ and the black hole parameters $\left(a,e,A,L\right)$ kept fixed, have positive imaginary part $\mathrm{Im}\,\omega_{0,\ell,0} > 0$. This suggests that the black holes with $A = 0.918$, $a = e = 0.0001$, $L = 1$, and $0.65 \leq r_{+} \leq 0.653$ are unstable (see Table \ref{table:4}). In addition, these frequencies are consistent with the photon sphere modes found in \cite{Xiong:2023usm}. It is worth mentioning that although at the spatial infinity the roots of the indicial equation differ by an integer, in contrast with the non-resonant spectra of the eigenvalue matrix $J_{\nu}$ in \eqref{eq:M_nu}, the isomonodromic approach remains valid. The trinion containing such singular point is reducible, while the other two remain generic with solutions given by the three-point Fuchsian system \eqref{eq:3point}. The latter configuration will simplify the Fredholm determinant representation of the tau function.

Despite the narrow space of black hole parameters that we have investigated in this work, the set of initial conditions solve the eigenvalue problem for ODEs with five regular singular points in general. The Heun-like equation \eqref{eq:heun-like_equation} appears in other problems, such as massive scalar and fermionic perturbations in Kerr-(A)dS$_{4}$ black holes \cite{Kraniotis:2016maw,Kraniotis:2018zmh}, and therefore, deserves further investigation. In addition, one can investigate physical quantities such as greybody factors and Love numbers in accelerating black holes by means of the connection coefficients of the Heun equations \cite{Bonelli:2021uvf,Bonelli:2022ten}. Finally, it would be interesting to consider the extremal limit for this configuration, as the resulting ODE will have three regular and one irregular singular point. To our knowledge, this has not been explored in the literature yet.

\begin{acknowledgments}
This research was supported by Basic Science Research Program through the National Research Foundation of Korea (NRF) funded by the Ministry of Education (NRF-2022R1I1A2063176) and the Dongguk University Research Fund of 2023. BG appreciates APCTP for its hospitality during the topical research program, {\it Multi-Messenger Astrophysics and Gravitation}. J.B.A. acknowledges Bruno Carneiro da Cunha and Jo\~{a}o Paulo Cavalcante for stimulating discussions about this work. Finally, we thank the referee for motivating a revision of the previous version of this paper.
\end{acknowledgments}

\appendix
\section{An alternative radial dictionary}
\label{appendix_A}
As we have discussed in Section \ref{sec:2.1.1}, in the conformally coupled case, the point $w = w_{1}$ is an apparent singularity that can be removed from the equation \eqref{eq:angular_odeS}. Likewise, the point $r = \infty$ can be extracted from the radial equation by setting the characteristic exponent $\beta_{\infty}^{\pm}$. 
To illustrate the fate of the singularity at $r = \infty$, let us consider the following change of variables
\begin{equation}\label{eq:dictionary_02}
z = \frac{r_{+} - r_{0}}{r_{+} - r_{1}}\frac{r - r_{1}}{r - r_{0}},
\end{equation}
with the conformal moduli given by
\begin{equation}\label{eq:rad_conformal_moduli_02}
z_{1} = \frac{r_{-} - r_{1}}{r_{-} - r_{0}}\frac{r_{+} - r_{0}}{r_{+} - r_{1}}, \qquad z_{\infty} = \frac{r_{+} - r_{0}}{r_{+} - r_{1}}.
\end{equation}
Furthermore, by introducing the following transformation
\begin{equation}\label{eq:shomotopic_radial_02}
R(z) = z^{\beta^{-}_{1}} (z - z_{1})^{\beta^{-}_{-}} (z - z_{\infty})^{\Delta/2} (z - 1)^{\beta^{-}_{+}} f(z),
\end{equation}
the radial ODE \eqref{eq:radial_ode} transforms into the canonical form of a Heun-like equation
\begin{equation}\label{eq:heun_like_with_delta}
\begin{split}
\frac{d^{2}f}{d z^{2}} &+ \biggl(\frac{1 - \theta_{1}}{z} + \frac{1 - \theta_{-}}{z - z_{1}} + \frac{\Delta - 2}{z - z_{\infty}} + \frac{1 - \theta_{+}}{z - 1} \biggr)\frac{d f}{d z} \\
&\qquad + \biggl[\frac{q}{z(z - 1)} - \frac{z_{1}(z_{1} - 1) K_{1}}{z(z - z_{1})(z - 1)}-\frac{z_{\infty}(z_{\infty} - 1) K_{\infty}}{z(z - z_{\infty})(z - 1)} + \frac{\frac{1}{4}\left(\Delta - 4\right)\left(\Delta - 2 \right)}{(z - z_{\infty})^{2}}\biggr]f(z) = 0,
\end{split}
\end{equation}
where
\begin{equation}
\begin{split}
&q = \frac{1}{4}\left[\left(\theta_{1} + \theta_{-} + \theta_{+} - \Delta\right)^{2} - \theta_{0}^{2} - \left(\Delta - 4\right)\left(\Delta - 2\right)\right],\\
&K_{1} = \frac{L^{2}\lambda}{\left(1 - A^{2}L^{2}\right)\left(r_{+} - r_{1}\right)\left(r_{-} - r_{0}\right) z_{1} (z_{1} - 1)} - \frac{1}{z_{1}}\biggl( \frac{1}{3} - \frac{1}{2}\left(\theta_{1} + \theta_{-}\right) \biggr) \\
&\qquad - \frac{1}{z_{1} - 1}\biggl( \frac{1}{3} - \frac{1}{2}\left(\theta_{-} + \theta_{+}\right) \biggr)
- \frac{1}{z_{1} - z_{\infty}}\frac{\left(\Delta - 2\right)\left(1 - \theta_{-}\right)}{2}, \\
&K_{\infty} = \frac{\left(\Delta - 2\right)}{2}\left[\frac{\theta_{1} - 1}{z_{\infty}} + \frac{\theta_{-} - 1}{z_{\infty} - z_{1}} + \frac{\theta_{+} - 1}{z_{\infty} - 1}\right].
\end{split}
\end{equation}
Then, one can see that by fixing $\Delta = 2$ (or $\beta^{-}_{\infty} = 1$), the singularity at $z = z_{\infty}$ can be removed from \eqref{eq:heun_like_with_delta}, yielding
\begin{equation}\label{eq:heun_minus}
\frac{d^{2}f}{d z^{2}} + \biggl(\frac{1 - \theta_{1}}{z} + \frac{1 - \theta_{-}}{z - z_{1}} + \frac{1 - \theta_{+}}{z - 1}\biggr)\frac{d f}{d z}
+ \biggl[\frac{\tilde{\kappa}_{1}\tilde{\kappa}_{2}}{z(z - 1)} - \frac{z_{1}(z_{1} - 1)\tilde{K}_{1}}{z(z - z_{1})(z - 1)}\biggr]f(z) = 0,
\end{equation}
with
\begin{subequations}
\begin{equation}\label{eq:kappas_minus}
\tilde{\kappa}_{1} = \dfrac{1}{2}\left(\theta_{1} + \theta_{-} + \theta_{+} - 2 + \theta_{0}\right), \qquad \tilde{\kappa}_{2} = \dfrac{1}{2}\left(\theta_{1} + \theta_{-} + \theta_{+} - 2 - \theta_{0}\right),
\end{equation}
\begin{equation}\label{eq:accessory_minus}
\tilde{K}_{1} = \frac{L^{2}\lambda}{\left(1 - A^{2}L^{2}\right)\left(r_{+} - r_{1}\right)\left(r_{-} - r_{0}\right) z_{1} (z_{1} - 1)} - \frac{1}{z_{1}}\biggl(\frac{1}{3} - \frac{1}{2}\left(\theta_{1} + \theta_{-}\right)\biggr) - \frac{1}{z_{1} - 1}\biggl(\frac{1}{3} - \frac{1}{2}\left(\theta_{-} + \theta_{+}\right)\biggr).
\end{equation}
\end{subequations}

\section{The isomonodromic deformation approach}
\label{appendix_B}

Let us study the $2 \times 2$ linear ODE system with five singularities
\begin{equation}\label{eq:fuchsian}
\dfrac{d\Phi}{dz} = \left[\frac{A_{0}}{z} + \frac{A_{1}}{z-t_{1}} + \frac{A_{2}}{z-t_{2}} + \frac{A_{3}}{z-1}\right]\Phi,
\end{equation}
where $\Phi(z)$ is the $2 \times 2$ fundamental matrix of solutions as \eqref{eq:Phi}, and the residue matrices $A_{i}, \left(i =  0,1,2,3 \right)$ do not depend on $z$. Furthermore, since the point at infinity is a regular singular point, we find that
\begin{equation}\label{eq:A_infty}
A_{\infty} = -\left(A_{0} + A_{1} + A_{2} + A_{3}\right),
\end{equation}
and, as a consequence of the isomonodromy condition, the matrix coefficients $A_{i}$ satisfy the Schlesinger system:
\begin{equation}\label{eq:schlesinger_five}
\frac{\partial A_{i}}{\partial t_{k}} = \frac{\left[A_{k},A_{i}\right]}{t_{k}-t_{i}}\quad \mathrm{for} \quad k \neq i, \quad \mathrm{and} \quad \frac{\partial A_{k}}{\partial t_{k}} = -\sum_{i \neq k}\frac{\left[A_{k},A_{i}\right]}{t_{k}-t_{i}},
\end{equation}
which implies that $A_{\infty} = \mathrm{const}$, and using the freedom of global conjugation, one may assume
\begin{equation}\label{eq:diag_A_infty}
A_{\infty} = \begin{pmatrix}
\kappa_{1} & 0 \\
0 & \kappa_{2} \\
\end{pmatrix}.
\end{equation}
Then the entry $\left[A(z)\right]_{12}$ of the matrix $A(z)$ is given by
\begin{equation}\label{eq:A_12}
\left[A(z)\right]_{12} = \frac{k(z-\lambda_{1})(z-\lambda_{2})}{z(z-t_{1})(z-t_{2})(z-1)}, 
\end{equation}
with some constant $k \in \mathds{C}$ and $\lbrace \lambda_{1}, \lambda_{2} \rbrace$ are a set of zeros in the complex plane.

The Fuchsian system \eqref{eq:fuchsian} has an associated second-order ordinary differential equation for the first row of the fundamental matrix
\begin{equation}\label{eq:fuchsian_ODE}
y^{\prime\prime}_{j}(z) - \left( \Tr A + \frac{\left[A^{\prime}(z)\right]_{12}}{\left[A(z)\right]_{12}} \right)y^{\prime}_{j}(z) + \left(\det A -\left[A^{\prime}(z)\right]_{11} + \left[A(z)\right]_{11}\frac{\left[A^{\prime}(z)\right]_{12}}{\left[A(z)\right]_{12}}\right)y_{j}(z) = 0, \quad \left( j = 1,2 \right),
\end{equation}
where the prime denotes a derivative with respect to $z$, and we will omit the subscript $j$ in $y_{j}$. By introducing \eqref{eq:gauge}, \eqref{eq:A_infty} and \eqref{eq:A_12}, the equation reduces to
\begin{equation}\label{eq:ODE}
y^{\prime\prime}(z) + p(z)y^{\prime} + q(z)y(z) = 0,
\end{equation}
with
\begin{equation}
p(z) = \frac{1-\Tr A_{0}}{z} + \frac{1-\Tr A_{1}}{z-t_{1}} + \frac{1-\Tr A_{2}}{z-t_{2}} + \frac{1-\Tr A_{3}}{z-1}-\frac{1}{z-\lambda_{1}}-\frac{1}{z-\lambda_{2}},
\end{equation}
and
\begin{equation}
\begin{gathered}
q(z) = \frac{\det A_{0}}{z^{2}} + \frac{\det A_{1}}{(z-t_{1})^{2}} + \frac{\det A_{2}}{(z-t_{2})^{2}} + \frac{\det A_{3}}{(z-1)^{2}} + \frac{\kappa}{z(z-1)} - \frac{t_{1}(t_{1}-1)\mathcal{K}_{1}}{z(z-t_{1})(z-1)} - \frac{t_{2}(t_{2}-1)\mathcal{K}_{2}}{z(z-t_{2})(z-1)} \\ 
+ \frac{\left[A_{\infty}\right]_{11}}{z(z-1)} + \frac{\lambda_{1}(\lambda_{1}-1)\mu_{1}}{z(z-\lambda_{1})(z-1)} +\frac{\lambda_{2}(\lambda_{2}-1)\mu_{2}}{z(z-\lambda_{2})(z-1)},
\end{gathered}
\end{equation}
where
\begin{subequations}
\begin{equation}
\kappa = \det A_{\infty} - \det A_{0} - \det A_{1} - \det A_{2} -\det A_{3},
\end{equation}
\begin{equation}
\mu_{1} = \frac{\left[A_{0}\right]_{11}}{\lambda_{1}} + \frac{\left[A_{1}\right]_{11}}{\lambda_{1}-t_{1}} + \frac{\left[A_{2}\right]_{11}}{\lambda_{1}-t_{2}}+\frac{\left[A_{3}\right]_{11}}{\lambda_{1}-1}, 
\end{equation}
\begin{equation}
\mu_{2} = \frac{\left[A_{0}\right]_{11}}{\lambda_{2}} + \frac{\left[A_{1}\right]_{11}}{\lambda_{2}-t_{1}} + \frac{\left[A_{2}\right]_{11}}{\lambda_{2}-t_{2}}+\frac{\left[A_{3}\right]_{11}}{\lambda_{2}-1}, 
\end{equation}
\begin{equation}\label{eq:Kone}
\begin{split}
\mathcal{K}_{1} = &\frac{1}{t_{1}}\Tr (A_{0}A_{1}) + \frac{1}{t_{1}-t_{2}}\Tr (A_{1}A_{2}) + \frac{1}{t_{1}-1}\Tr (A_{1}A_{3})
- \frac{1}{t_{1}}\Tr A_{0}\Tr A_{1} - \frac{1}{t_{1}-t_{2}}\Tr A_{1}\Tr A_{2} \\
&-\frac{1}{t_{1}-1}\Tr A_{1}\Tr A_{3} + \frac{\left[A_{0}\right]_{11}+\left[A_{1}\right]_{11}}{t_{1}} + \frac{\left[A_{1}\right]_{11}+\left[A_{2}\right]_{11}}{t_{1}-t_{2}} + \frac{\left[A_{1}\right]_{11}+\left[A_{3}\right]_{11}}{t_{1}-1}-\frac{\left[A_{1}\right]_{11}}{t_{1}-\lambda_{1}} - \frac{\left[A_{1}\right]_{11}}{t_{1}-\lambda_{2}},
\end{split}
\end{equation}
\begin{equation}\label{eq:Ktwo}
\begin{split}
\mathcal{K}_{2} = &\frac{1}{t_{2}}\Tr (A_{0}A_{2}) + \frac{1}{t_{2}-t_{1}}\Tr (A_{2}A_{1}) + \frac{1}{t_{2}-1}\Tr (A_{2}A_{3})
- \frac{1}{t_{2}}\Tr A_{0}\Tr A_{2} - \frac{1}{t_{2}-t_{1}}\Tr A_{2}\Tr A_{1} \\
&-\frac{1}{t_{2}-1}\Tr A_{2}\Tr A_{3} + \frac{\left[A_{0}\right]_{11}+\left[A_{2}\right]_{11}}{t_{2}} + \frac{\left[A_{1}\right]_{11}+\left[A_{2}\right]_{11}}{t_{2}-t_{1}} + \frac{\left[A_{2}\right]_{11}+\left[A_{3}\right]_{11}}{t_{2}-1}-\frac{\left[A_{2}\right]_{11}}{t_{2}-\lambda_{1}} - \frac{\left[A_{2}\right]_{11}}{t_{2}-\lambda_{2}}.
\end{split}
\end{equation}
\end{subequations}
In addition, we choose the following properties for the residue matrices
\begin{equation}\label{eq:matrices}
\Tr {A_{i}} = \vartheta_{i}, \qquad \det A_{i} = 0, \qquad \left( i = 0,1,2,3 \right),
\end{equation}
which bring \eqref{eq:ODE} into the form
\begin{equation}\label{eq:deformed_five}
\begin{split}
y^{\prime\prime}(z) + \biggl[\frac{1-\vartheta_{0}}{z} &+ \frac{1-\vartheta_{1}}{z-t_{1}} + \frac{1-\vartheta_{2}}{z-t_{2}} + \frac{1-\vartheta_{3}}{z-1} - \frac{1}{z-\lambda_{1}} - \frac{1}{z-\lambda_{2}}\biggr]y^{\prime}(z) \\
&+ \biggl(\frac{\kappa_{1}(1+\kappa_{2})}{z(z-1)} - \frac{t_{1}(t_{1}-1)\mathcal{K}_{1}}{z(z-t_{1})(z-1)} - \frac{t_{2}(t_{2}-1)\mathcal{K}_{2}}{z(z-t_{2})(z-1)} \\
&\qquad\qquad\qquad+ \frac{\lambda_{1}(\lambda_{1}-1)\mu_{1}}{z(z-\lambda_{1})(z-1)} + \frac{\lambda_{2}(\lambda_{2}-1)\mu_{2}}{z(z-\lambda_{2})(z-1)}\biggr)y(z) = 0.
\end{split}
\end{equation}
One can show that the singular points at $z = \lambda_{j}$ are apparent singularities, as the characteristic exponents are integers, and therefore the monodromies around these points are trivial. The latter impose an algebraic relation between $\mathcal{K}_{j}$, $\mu_{j}$, $\lambda_{j}$ and $t_{j}$,
\begin{subequations}\label{eq:K1_and_K2}
\begin{equation}
\begin{split}
\mathcal{K}_{1} = &- \frac{\lambda_{1}(\lambda_{1}-t_{1})(\lambda_{1}-t_{2})(\lambda_{1}-1)(\lambda_{2}-t_{1})}{t_{1}(t_{1}-t_{2})(t_{1}-1)(\lambda_{1}-\lambda_{2})}\biggl[\mu_{1}^{2} - \left(\frac{\vartheta_{0}}{\lambda_{1}} + \frac{\vartheta_{1}-1}{\lambda_{1}-t_{1}}+\frac{\vartheta_{2}}{\lambda_{1}-t_{2}}+\frac{\vartheta_{3}}{\lambda_{1}-1}\right)\mu_{1} \\ 
&+ \frac{\kappa_{1}(1+\kappa_{2})(\lambda_{1}-\lambda_{2})}{\lambda_{1}(\lambda_{1}-t_{2})(\lambda_{1}-1)}\biggr] -\frac{\lambda_{2}(\lambda_{2}-t_{1})(\lambda_{2}-t_{2})(\lambda_{2}-1)(\lambda_{1}-t_{1})}{t_{1}(t_{1}-t_{2})(t_{1}-1)(\lambda_{2}-\lambda_{1})}\biggl[\mu_{2}^{2} \\
&- \left(\frac{\vartheta_{0}}{\lambda_{2}} + \frac{\vartheta_{1}-1}{\lambda_{2}-t_{1}}+\frac{\vartheta_{2}}{\lambda_{2}-t_{2}}+\frac{\vartheta_{3}}{\lambda_{2}-1}\right)\mu_{2}\biggr],
\end{split}
\end{equation}
\begin{equation}
\begin{split}
\mathcal{K}_{2} = &- \frac{\lambda_{2}(\lambda_{2}-t_{1})(\lambda_{2}-t_{2})(\lambda_{2}-1)(\lambda_{1}-t_{2})}{t_{2}(t_{2}-t_{1})(t_{2}-1)(\lambda_{2}-\lambda_{1})}\biggl[\mu_{2}^{2} - \left(\frac{\vartheta_{0}}{\lambda_{2}} + \frac{\vartheta_{1}}{\lambda_{2}-t_{1}}+\frac{\vartheta_{2}-1}{\lambda_{2}-t_{2}}+\frac{\vartheta_{3}}{\lambda_{2}-1}\right)\mu_{2} \\ 
&+ \frac{\kappa_{1}(1+\kappa_{2})(\lambda_{2}-\lambda_{1})}{\lambda_{2}(\lambda_{2}-t_{1})(\lambda_{2}-1)}\biggr] -\frac{\lambda_{1}(\lambda_{1}-t_{1})(\lambda_{1}-t_{2})(\lambda_{1}-1)(\lambda_{2}-t_{2})}{t_{2}(t_{2}-t_{1})(t_{2}-1)(\lambda_{1}-\lambda_{2})}\biggl[\mu_{1}^{2} \\
&- \left(\frac{\vartheta_{0}}{\lambda_{1}} + \frac{\vartheta_{1}}{\lambda_{1}-t_{1}}+\frac{\vartheta_{2}-1}{\lambda_{1}-t_{2}}+\frac{\vartheta_{3}}{\lambda_{1}-1}\right)\mu_{1}\biggr].
\end{split}
\end{equation}
\end{subequations}
Okamoto showed that the isomonodromic deformation equations \eqref{eq:schlesinger_five} of the Fuchsian system \eqref{eq:fuchsian} can be obtained from the Hamiltonian system \cite{okamoto1981isomonodromic}:
\begin{equation}\label{eq:Hamilton_eqns}
\frac{\partial \lambda_{k}}{\partial t_{j}} = \frac{\partial \mathcal{K}_{j}}{\partial \mu_{k}}, \qquad \frac{\partial \mu_{k}}{\partial t_{j}} = -\frac{\partial \mathcal{K}_{j}}{\partial \lambda_{k}}, \qquad \left(j,k= 1,2 \right),
\end{equation}
where $(\lambda_{j},\mu_{j}),$ are canonical variables, $\mathcal{K}_{j}$ is the Hamiltonian. Such dynamical system \eqref{eq:Hamilton_eqns} is known as the two-dimensional Garnier system, denoted by $G_{2}(\vartheta), \,\,\, \vartheta = \lbrace \vartheta_{0},\vartheta_{1},\vartheta_{2},\vartheta_{3},\vartheta_{\infty} \rbrace$, and is completely integrable. Furthermore, we can introduce an isomonodromic tau function \`{a} la Jimbo-Miwa-Ueno (JMU), defined by 
\begin{equation}\label{eq:tau_function}
\frac{\partial}{\partial t_{k}}\log \tau_{\mathrm{JMU}} = \sum_{\ell=0,\ell \neq k}^{3}\frac{1}{t_{k} - t_{\ell}}\Tr A_{k}A_{\ell}, \quad\quad \left( k=  1,2 \right),
\end{equation}
which is a closed $1$-form provided that the Schlesinger equations \eqref{eq:schlesinger_five} are satisfied. The definition \eqref{eq:tau_function} assumes that residue matrices are traceless $A_{i} \in \mathfrak{sl}(2,\mathbb{C})$, which is in contrast with our choice \eqref{eq:matrices}. In principle those residue matrices will correspond to a different Fuchsian system \eqref{eq:fuchsian}, but one can check that the by performing the following transformation $A^{\prime}_{i} = A_{i} - \tfrac{1}{2}\vartheta_{i}\mathds{1}$, the solutions of Fuchsian systems of traceless and nonzero trace residue matrices are related by multiplicative factors. Namely, the isomonodromic tau functions are related by
\begin{equation}\label{eq:gauge}
\frac{\partial }{\partial t_{k}}\log \tau^{\prime}_{\mathrm{JMU}} = \frac{\partial }{\partial t_{k}}\log \tau_{\mathrm{JMU}}-\frac{1}{2}\sum_{\ell=0,\ell \neq k}^{3}\frac{\vartheta_{k}\vartheta_{\ell}}{t_{k}-t_{\ell}}, \quad\quad \left( k= 1,2 \right).
\end{equation}
Thus by replacing \eqref{eq:gauge} into \cref{eq:Kone,eq:Ktwo}, as well as equating to their corresponding Hamiltoninian in \eqref{eq:K1_and_K2}, we obtain
\begin{subequations}\label{eq:log_derivatives}
\begin{equation}\label{eq:log_tau_t1}
\begin{split}
\frac{\partial}{\partial t_{1}}\log \tau^{\prime}_{\mathrm{JMU}} = 
&\frac{\vartheta_{0}\vartheta_{1}}{2 t_{1}} + \frac{\vartheta_{1}\vartheta_{2}}{2(t_{1}-t_{2})}+\frac{\vartheta_{1}\vartheta_{3}}{2(t_{1}-1)} - \frac{\left[A_{0}\right]_{11}+\left[A_{1}\right]_{11}}{t_{1}} \\
&- \frac{\left[A_{1}\right]_{11}+\left[A_{2}\right]_{11}}{t_{1}-t_{2}} - \frac{\left[A_{1}\right]_{11}+\left[A_{3}\right]_{11}}{t_{1}-1} +\frac{\left[A_{1}\right]_{11}}{t_{1}-\lambda_{1}} + \frac{\left[A_{1}\right]_{11}}{t_{1}-\lambda_{2}} + \mathcal{K}_{1},
\end{split}
\end{equation}
\begin{equation}\label{eq:log_tau_t2}
\begin{split}
\frac{\partial}{\partial t_{2}}\log \tau^{\prime}_{\mathrm{JMU}} = 
&\frac{\vartheta_{0}\vartheta_{2}}{2 t_{2}} + \frac{\vartheta_{2}\vartheta_{1}}{2(t_{2}-t_{1})}
+\frac{\vartheta_{2}\vartheta_{3}}{2(t_{2}-1)} - \frac{\left[A_{0}\right]_{11}+\left[A_{2}\right]_{11}}{t_{2}} \\
&- \frac{\left[A_{1}\right]_{11}+\left[A_{2}\right]_{11}}{t_{2}-t_{1}} - \frac{\left[A_{2}\right]_{11}+\left[A_{3}\right]_{11}}{t_{2}-1}+\frac{\left[A_{2}\right]_{11}}{t_{2}-\lambda_{1}} + \frac{\left[A_{2}\right]_{11}}{t_{2}-\lambda_{2}} + \mathcal{K}_{2}.
\end{split}
\end{equation}
\end{subequations}
Given the isomonodromic tau function for \eqref{eq:log_derivatives}, we can establish the Riemann-Hilbert map to relate the accessory parameters of the Fuchsian ODEs to the monodromy data. One can show that by setting initial conditions to the Hamiltonian system, the deformed equation \eqref{eq:deformed_five} recovers a Heun-like equation with five singularities
of the form
\begin{equation}\label{eq:heun-like_equation}
\begin{split}
\frac{d^{2}y}{d z^{2}} + \biggl(\frac{1-\theta_{0}}{z}+\frac{1-\theta_{1}}{z-z_{1}} &+ \frac{1-\theta_{2}}{z-z_{2}}+\frac{1-\theta_{3}}{z-1}\biggr)\frac{d y}{d z}\\
&+\biggl[\frac{\kappa_{+}\kappa_{-}}{z(z-1)} - \frac{z_{1}(z_{1}-1)K_{1}}{z(z-z_{1})(z-1)} - \frac{z_{2}(z_{2}-1)K_{2}}{z(z-z_{2})(z-1)}\biggr]y(z) = 0,
\end{split}
\end{equation}
where each apparent singularity has merged with one regular singular point. Namely, we require the following initial conditions
\begin{equation}\label{eq:initial_conditions}
\begin{split}
\lambda_{1}(t_{1}=z_{1},t_{2}=z_{2}) = z_{1}, \quad \mu_{1}(t_1=z_{1},t_2=z_{2}) = \frac{K_{1}}{1-\theta_{1}},\\
\lambda_{2}(t_{1}=z_{1},t_{2}=z_{2}) = z_{2}, \quad \mu_{2}(t_1=z_{1},t_2=z_{2}) = \frac{K_{2}}{1-\theta_{2}},
\end{split}
\end{equation}
while the local monodromies will change as follows 
\begin{equation}\label{eq:thetas_x_varthetas}
\vartheta_{0} = \theta_{0}, \quad \vartheta_{1} = \theta_{1}-1, \quad \vartheta_{2} = \theta_{2}-1, \quad \vartheta_{3} = \theta_{3}, \quad \vartheta_{\infty} = \theta_{\infty}+1.
\end{equation}
The conditions \eqref{eq:initial_conditions} yield a well-posed initial value problem for the Garnier system \eqref{eq:Hamilton_eqns}. As it has been shown in \cite{CarneirodaCunha:2021jsu}, these conditions can be rewritten in terms of the isomonodromic tau function. In turn, the initial conditions for $\left(\mu_{k}(\lbrace z_{k} \rbrace),\lambda_{k}(\lbrace z_{k} \rbrace)\right)$ will be related to the logarithmic derivative and a zero of the tau function $\tau_{\rm JMU}$ with given monodromy data. Namely, we have
\begin{subequations}\label{eq:initial_value_tau}
\begin{equation}\label{eq:dlog_tau}
\frac{\partial}{\partial t_{k}}\log \tau_{\mathrm{JMU}}\left(\rho;\lbrace z_{k}\rbrace \right) = K_{k} 
+ \frac{\vartheta_{0}\vartheta_{k}}{2 t_{k}} + \sum^{2}_{l \neq k}\frac{\vartheta_{k}\vartheta_{l}}{2(t_{k}-t_{l})}+\frac{\vartheta_{k}\vartheta_{3}}{2(t_{k}-1)}, \qquad \left( k = 1,2 \right),
\end{equation}
\begin{equation}\label{eq:zero_tau}
\tau_{\rm JMU}\left(\rho^{+}_{k};\lbrace z_{k} \rbrace\right) = 0,
\end{equation}
\end{subequations}
where the monodromy arguments $\rho$ and $\rho_{k}^{+}$ are related by shifts:
\begin{subequations}\label{eq:monodromy_data}
\begin{equation}\label{eq:data0}
\rho = \lbrace \vartheta;\sigma_{k};\eta_{k} \rbrace = \lbrace \theta_{0}, \theta_{1} - 1, \theta_{2} - 1, \theta_{3}, \theta_{\infty} + 1; \sigma_{1} - 1, \sigma_{2} - 1; \eta_{1}, \eta_{2} \rbrace,
\end{equation}
\begin{equation}\label{eq:data1}
\rho_{1}^{+} = \lbrace \vartheta^{+};\sigma^{+}_{k};\eta^{+}_{k} \rbrace = \lbrace \theta_{0}, \theta_{1}, \theta_{2} - 1, \theta_{3}, \theta_{\infty}; \sigma_{1}, \sigma_{2}; \eta_{1}, \eta_{2} \rbrace,
\end{equation}
\begin{equation}\label{eq:data2}
\rho_{2}^{+} = \lbrace \vartheta^{+};\sigma^{+}_{k};\eta^{+}_{k} \rbrace = \lbrace \theta_{0}, \theta_{1} - 1, \theta_{2}, \theta_{3}, \theta_{\infty}; \sigma_{1} - 1, \sigma_{2}; \eta_{1}, \eta_{2} \rbrace,
\end{equation}
\end{subequations}
where the set of transcendental equations \eqref{eq:initial_value_tau} supplemented with \eqref{eq:monodromy_data} will solve the boundary value problem for the radial ODE \eqref{eq:heun_radial}.

\section{The tau function for Fuchsian systems with five regular singular points}
\label{appendix_C}

The Jimbo-Miwa-Ueno isomonodromic tau function $\tau_{\rm JMU}(t)$ admits a block determinant representation
\begin{equation}\label{eq:tau_jmu}
\tau_{\rm JMU}(t) = \Upsilon (t)\cdot \det \left(\mathds{1}-\mathrm{K}\right),
\end{equation}
where the operator $\mathrm{K}$ is
\begin{equation}\label{eq:fredholm}
\mathrm{K} = \begin{pmatrix}
    0 & \mathsf{a}^{[2]} & \mathsf{b}^{[2]} & 0 & 0 & 0 & \ldots & 0 & 0 \\
    \mathsf{d}^{[1]} & 0 & 0 & 0 & 0 & 0 & \dots & 0 & 0 \\
    0 & 0 & 0 & \mathsf{a}^{[3]} & \mathsf{b}^{[3]} & 0 & \ldots & 0 & 0 \\
    0 & \mathsf{c}^{[2]} & \mathsf{d}^{[2]} & 0 & 0 & 0 & \ldots & 0 & 0 \\
    0 & 0 & 0 & 0 & 0 & \mathsf{a}^{[4]} & \ldots & 0 & 0 \\
    0 & 0 & 0 & \mathsf{c}^{[3]} & \mathsf{d}^{[3]} & 0 & \ldots & 0 & 0 \\
    \vdots & \vdots & \vdots & \vdots & \vdots & \vdots & \ddots & \vdots & \vdots \\
    0 & 0 & 0 & 0 & 0 & 0 & \ldots & 0 & \mathsf{a}^{[n-2]} \\
    0 & 0 & 0 & 0 & 0 & 0 & \ldots & \mathsf{d}^{[n-3]} & 0\\
\end{pmatrix},
\end{equation}
$n$ corresponds to the number of regular singular points on $\mathds{P}^{1}$, and $\Upsilon (t)$ is an analytic function of the positions of the remaining singular points that cannot be fixed by a M\"{o}bius transformation  $t = \lbrace t_{1},\ldots, t_{n-3}\rbrace$. 

The operators $\mathsf{a}^{[k]},\mathsf{b}^{[k]},\mathsf{c}^{[k]},\mathsf{d}^{[k]}$ are Cauchy-Plemelj operators $\mathcal{P}_{\bigoplus}:\mathcal{H}\rightarrow \mathcal{H}$ of the form:
\begin{equation}
\begin{split}
\left(\mathsf{a}^{[k]}g\right)(z)=\oint_{\mathcal{C}^{[k-1]}}\frac{dz^{\prime}}{2\pi i}\mathsf{a}^{[k]}(z,z^{\prime})g(z^{\prime}), \qquad \left(\mathsf{b}^{[k]}g\right)(z)=\oint_{\mathcal{C}^{[k]}}\frac{dz^{\prime}}{2\pi i}\mathsf{b}^{[k]}(z,z^{\prime})g(z^{\prime}), \\
\left(\mathsf{c}^{[k]}g\right)(z)=\oint_{\mathcal{C}^{[k-1]}}\frac{dz^{\prime}}{2\pi i}\mathsf{c}^{[k]}(z,z^{\prime})g(z^{\prime}), \qquad \left(\mathsf{d}^{[k]}g\right)(z)=\oint_{\mathcal{C}^{[k]}}\frac{dz^{\prime}}{2\pi i}\mathsf{d}^{[k]}(z,z^{\prime})g(z^{\prime}),
\end{split}
\end{equation}
where the integral kernels of the three-point projection operators can be expanded in a Fourier basis converging inside the trinion $\mathcal{T}^{[k]}$,
\begin{subequations}\label{eq:Fourier}
\begin{equation}
\mathsf{a}^{[k]}(z,z^{\prime}) = \frac{\Psi^{[k]}_{+}(z)\Psi^{[k]}_{+}(z^{\prime})^{-1}-\mathds{1}}{z-z^{\prime}} = \sum_{p,q = 1}^{\infty}\left(\mathsf{A}^{[k]}\right)^{p}_{q}z^{p-1}z^{\prime}\,^{q-1}, \qquad z \in \mathcal{C}^{[k-1]},
\end{equation}
\begin{equation}
\mathsf{b}^{[k]}(z,z^{\prime}) = -\frac{\Psi^{[k]}_{+}(z)\Psi^{[k]}_{+}(z^{\prime})^{-1}}{z-z^{\prime}}=\sum_{p,q = 1}^{\infty}\left(\mathsf{B}^{[k]}\right)^{p}_{q}z^{p-1}z^{\prime}\,^{-q}, \qquad z \in \mathcal{C}^{[k-1]},
\end{equation}
\begin{equation}
\mathsf{c}^{[k]}(z,z^{\prime}) = \frac{\Psi^{[k]}_{+}(z)\Psi^{[k]}_{+}(z^{\prime})^{-1}}{z-z^{\prime}}=\sum_{p,q = 1}^{\infty}\left(\mathsf{C}^{[k]}\right)^{p}_{q}z^{-p}z^{\prime}\,^{q-1}, \qquad z \in \mathcal{C}^{[k]},
\end{equation}
\begin{equation}
\mathsf{d}^{[k]}(z,z^{\prime}) = \frac{\mathds{1}-\Psi^{[k]}_{+}(z)\Psi^{[k]}_{+}(z^{\prime})^{-1}}{z-z^{\prime}}=\sum_{p,q = 1}^{\infty}\left(\mathsf{D}^{[k]}\right)^{p}_{q}z^{-p}z^{\prime}\,^{-q}, \qquad z \in \mathcal{C}^{[k]},
\end{equation}
\end{subequations}
with the Fourier coefficients \eqref{eq:Fourier} given by semi-infinite matrices whose $N \times N$ blocks are determined by 
\begin{subequations}\label{eq:Fourier_coefficients}
\begin{equation}
\left(\mathsf{A}^{[k]}\right)^{p}_{q} = t_{k}^{\frac{1}{2}\sigma_{k-1}\bm{\sigma}_{3}}\left(\sum_{r=1}^{p}\left[\Psi^{[k]}\,^{-1}\right]_{q+r-1}\left[\Psi^{[k]}\right]_{p-r}t_{k}^{-p-q+1}\right)t_{k}^{-\frac{1}{2}\sigma_{k-1}\bm{\sigma}_{3}},
\end{equation}
\begin{equation}
\left(\mathsf{B}^{[k]}\right)^{p}_{q} =  \kappa_{k}^{\frac{1}{2}\bm{\sigma}_{3}}t_{k}^{\frac{1}{2}\sigma_{k}\bm{\sigma}_{3}}\left(\sum_{r=1}^{\mathrm{min}(p,q)}\left[\tilde{\Psi}^{[k]}\,^{-1}\right]_{q-r}\bm{\mathrm{G}}_{k}^{-1}\left[\Psi^{[k]}\right]_{p-r}t_{k}^{-p+q}\right)t_{k}^{-\frac{1}{2}\sigma_{k-1}\bm{\sigma}_{3}},
\end{equation}
\begin{equation}
\left(\mathsf{C}^{[k]}\right)^{p}_{q} = t_{k}^{\frac{1}{2}\sigma_{k}\bm{\sigma}_{3}}\left(\sum_{r=1}^{\mathrm{min}(p,q)}\left[\Psi^{[k]}\,^{-1}\right]_{q-r}\bm{\mathrm{G}}_{k}\left[\tilde{\Psi}^{[k]}\right]_{p-r}t_{k}^{p-q}\right)t_{k}^{\frac{1}{2}\sigma_{k-1}\bm{\sigma}_{3}}\kappa_{k}^{\frac{1}{2}\bm{\sigma}_{3}},
\end{equation}
\begin{equation}
\left(\mathsf{D}^{[k]}\right)^{p}_{q} = \kappa_{k}^{\frac{1}{2}\bm{\sigma}_{3}}t_{k}^{\frac{1}{2}\sigma_{k}\bm{\sigma}_{3}}\left(\sum_{r=1}^{p}\left[\tilde{\Psi}^{[k]}\,^{-1}\right]_{q+r-1}\left[\tilde{\Psi}^{[k]}\right]_{p-r}t_{k}^{p+q-1}\right)t_{k}^{-\frac{1}{2}\sigma_{k}\bm{\sigma}_{3}}\kappa_{k}^{-\frac{1}{2}\bm{\sigma}_{3}}.
\end{equation}
\end{subequations}
where the matrix functions $\Psi^{[k]}$ and $\tilde{\Psi}^{[k]}$ solve the Riemann-Hilbert Problem associated to $3$-point Fuchsian systems with prescribed monodromy. For $\vert t_{1} \vert \ll \vert t_{2} \vert \ll \cdots \ll 1$, the tau function can be expanded by truncating the kernels in the expansion of $\mathsf{a}^{[k]},\mathsf{b}^{[k]},\mathsf{c}^{[k]},\mathsf{d}^{[k]}$ up to order $Q \in \mathds{Z}_{> 0}$, such that the Fourier coefficients \eqref{eq:Fourier_coefficients} satisfy $p + q \leq Q$, denoted by $N Q \times N Q$-dimensional blocks (see \cite{Gavrylenko:2016zlf} for an explicit representation of the matrices). The latter gives the asymptotic expansion of $\tau_{\rm JMU}(t)$ to arbitrary finite order $Q$. Now, let us expand the parametrices
\begin{equation}
\begin{split}
&\Psi(\sigma_{k-1},\vartheta_{k},\sigma_{k};z/t_{k}) = \sum_{n=0}^{\infty}\left[\Psi^{[k]}\right]_{n}\left(\frac{z}{t_{k}}\right)^{n}, \\ 
&\Psi(-\sigma_{k},\vartheta_{k},\sigma_{k-1};t_{k}/z) = \sum_{n=0}^{\infty}\left[\tilde{\Psi}^{[k]}\right]_{n}\left(\frac{t_{k}}{z}\right)^{n},
\end{split}
\end{equation}
with $\bigl[\Psi^{[k]}\bigr]_{0} = \bigl[\tilde{\Psi}^{[k]}\bigr]_{0} = \mathds{1}$. In particular, for $N = 2$, $3$-point RHPs can be solved in terms of Gauss hypergeometric functions 
\begin{equation}\label{eq:3point}
\Psi(\alpha_{1},\alpha_{2},\alpha_{3};z) = \begin{pmatrix}
	\phi(\alpha_{1},\alpha_{2},\alpha_{3};z) & \chi(\alpha_{1},\alpha_{2},\alpha_{3};z) \\
	\chi(-\alpha_{1},\alpha_{2},\alpha_{3};z) & \phi(-\alpha_{1},\alpha_{2},\alpha_{3};z) \\
\end{pmatrix},
\end{equation}
where
\begin{equation}
\begin{split}
\phi(\alpha_{1},\alpha_{2},\alpha_{3};z) &= \,_{2}F_{1}\left(\frac{1}{2}(\alpha_{1}-\alpha_{2}+\alpha_{3}),\frac{1}{2}(\alpha_{1}-\alpha_{2}-\alpha_{3}),\alpha_{1};z\right), \\
\chi(\alpha_{1},\alpha_{2},\alpha_{3};z) &= \frac{\alpha_{3}^{2}-(\alpha_{2}-\alpha_{1})^{2}}{4\alpha_{1}(\alpha_{1}+1)}z\,_{2}F_{1}\left(1+\frac{1}{2}(\alpha_{1}-\alpha_{2}+\alpha_{3}),1+\frac{1}{2}(\alpha_{1}-\alpha_{2}-\alpha_{3}),2+\alpha_{1};z\right)
\end{split}
\end{equation}
and $\bm{\mathrm{G}}_{k}$ and $\kappa_{k}$ are given by
\begin{equation}
\bm{\mathrm{G}}_{k} = \frac{1}{2\sigma_{k}}\begin{pmatrix}
	\sigma_{k-1}+\vartheta_{k}+\sigma_{k} & \sigma_{k-1}+\vartheta_{k}-\sigma_{k} \\
	\sigma_{k-1}-\vartheta_{k}-\sigma_{k} & \sigma_{k-1}-\vartheta_{k}+\sigma_{k} \\
\end{pmatrix},
\end{equation}
\begin{equation}
\begin{split}
\kappa_{k} = &\,e^{i\eta_{k}}\frac{\Gamma^{2}\left(1-\sigma_{k}\right)}{\Gamma^{2}\left(1+\sigma_{k}\right)}\frac{\Gamma\left(1+\frac{1}{2}(\vartheta_{k}+\sigma_{k-1}+\sigma_{k})\right)\Gamma\left(1+\frac{1}{2}(\vartheta_{k}-\sigma_{k-1}+\sigma_{k})\right)}{\Gamma\left(1+\frac{1}{2}(\vartheta_{k}+\sigma_{k-1}-\sigma_{k})\right)\Gamma\left(1+\frac{1}{2}(\vartheta_{k}-\sigma_{k-1}-\sigma_{k})\right)}\\
&\times\frac{\Gamma\left(1+\frac{1}{2}(\vartheta_{k+1}+\sigma_{k+1}+\sigma_{k})\right)\Gamma\left(1+\frac{1}{2}(\vartheta_{k+1}-\sigma_{k+1}+\sigma_{k})\right)}{\Gamma\left(1+\frac{1}{2}(\vartheta_{k+1}+\sigma_{k+1}-\sigma_{k})\right)\Gamma\left(1+\frac{1}{2}(\vartheta_{k+1}-\sigma_{k+1}-\sigma_{k})\right)}.
\end{split}
\end{equation}
Furthermore, for $n=5$, the determinant reduces to
\begin{equation}\label{eq:operator_Kfive}
\mathrm{K} = \begin{pmatrix}
    0 & \mathsf{a}^{[2]} & \mathsf{b}^{[2]} & 0 \\
    \mathsf{d}^{[1]} & 0 & 0 & 0 \\
    0 & 0 & 0 & \mathsf{a}^{[3]} \\
    0 & \mathsf{c}^{[2]} & \mathsf{d}^{[2]} & 0
\end{pmatrix},
\end{equation}
satisfying the decomposition of the five-punctured Riemann sphere into three trinions:
\begin{equation}\label{eq:trinions}
\begin{split}
&\mathcal{T}^{[1]}: \sigma_{0} = \vartheta_{0}, \vartheta_{1}, \sigma_{1}, \eta_{1}, t_{1}; \\
&\mathcal{T}^{[2]}: \sigma_{1}, \vartheta_{2}, \sigma_{2}, \eta_{2}, t_{2}; \\ 
&\mathcal{T}^{[3]}: \sigma_{2}, \vartheta_{3}, \sigma_{3} = \vartheta_{\infty}, \eta_{3}=1, t_{3}=1,
\end{split}
\end{equation}
and the isomonodromic tau function reads
\begin{equation}\label{eq:tau_five}
\begin{split}
\tau_{\rm JMU}(\rho;t_{1},t_{2}) &= t_{1}^{\frac{1}{4}\left(\sigma_{1}^{2}-\vartheta_{0}^{2}-\vartheta_{1}^{2}\right)}t_{2}^{\frac{1}{4}\left(\sigma_{2}^{2}-\vartheta_{2}^{2}-\sigma_{1}^{2}\right)}\left(1-t_{1}\right)^{-\frac{1}{2}\vartheta_{1}\vartheta_{3}}\left(1-t_{2}\right)^{-\frac{1}{2}\vartheta_{2}\vartheta_{3}}\\
&\quad\times\left(1-\frac{t_{1}}{t_{2}}\right)^{-\frac{1}{2}\vartheta_{1}\vartheta_{2}}\det\left(1 - \mathsf{a}^{[2]}\mathsf{d}^{[1]}\right)\det \left(1 - \mathsf{d}^{[2]}\mathsf{a}^{[3]} - \mathsf{c}^{[2]}\mathsf{d}^{[1]}\left(1 - \mathsf{a}^{[2]}\mathsf{d}^{[1]}\right)^{-1}\mathsf{b}^{[2]}\mathsf{a}^{[3]}\right),
\end{split}
\end{equation}
where $t_{1} \ll t_{2} \ll 1$. The last expression will be introduced into the initial conditions \eqref{eq:initial_value_tau}, using the monodromy data \eqref{eq:monodromy_data}.

\bibliography{references}

\end{document}